\documentclass[pre,aps,twocolumn,superscriptaddress]{revtex4}

\usepackage{graphicx}
\usepackage{amsmath,empheq}
\usepackage{amssymb}
\usepackage{ifthen}
\usepackage{booktabs}

\usepackage{graphicx}
\usepackage{dcolumn}
\usepackage{bm}
\usepackage{gensymb}
\usepackage{color}

\newcommand{\pdiff}[3][]{\frac{\partial^{#1}#2}{\partial #3^{#1}}}

\makeatletter
\def\@frameeq#1{%
  \framebox{$\,\displaystyle#1\hbox{\vrule height 2.4ex depth 1.4ex width 0pt}\,$}}
\newcommand\Equation[1]{$$\refstepcounter{equation}%
  \@frameeq{#1}%
  \eqno \hbox{\@eqnnum}$$\@ignoretrue\ignorespaces}
\newcommand\Displaystyle[1]{$$\@frameeq{#1}$$\@ignoretrue\ignorespaces}
\makeatother

\usepackage{longtable}

\newcounter{subfigcount}
\newcounter{figcount}
\newcommand{\subfloat}[3]{%
{\ifnum\thefigure=\thefigcount\stepcounter{subfigcount}%
\else\setcounter{figcount}{\thefigure}\setcounter{subfigcount}{1}\fi%
}%
\noindent%
\begin{minipage}[b]{#1}%
  \centering%
  {#3}\\[0pt]%
  (\alph{subfigcount})~#2%
\end{minipage}}%

\newcommand{\subfloatflex}[2]{%
{\ifnum\thefigure=\thefigcount\stepcounter{subfigcount}
\else\setcounter{figcount}{\thefigure}\setcounter{subfigcount}{1}\fi%
}%
\noindent%
\begin{minipage}[b]{\widthof{#2}}%
  \centering%
  {#2}\\[0pt]%
  (\alph{subfigcount})~#1%
\end{minipage}}%

\begin{document}

\normalsize

\title{Three types of condensation in open wedges}

\author{Ji\v r\'\i \hspace{0.001cm} \surname{Janek}}
\affiliation{Research group of Molecular and Mesoscopic Modelling, The Czech Academy of Sciences, Institute of Chemical Process Fundamentals, 165 02 Prague, Czech Republic}
\author{Alexandr \surname{Malijevsk\'y}}
\affiliation{Research group of Molecular and Mesoscopic Modelling, The Czech Academy of Sciences, Institute of Chemical Process Fundamentals, 165 02 Prague, Czech
  Republic}
\affiliation{Department of Physical Chemistry, University of Chemistry and Technology Prague, 166 28 Prague, Czech Republic}

\begin{abstract}
  \noindent Condensation in linear wedges formed by semi-infinite walls is a well-established critical phenomenon characterized by the continuous
  growth of an adsorbed liquid layer as bulk two-phase coexistence is approached.  In this study, we investigate
  condensation in finite-length wedges open at both ends, demonstrating that the process becomes first-order. The open boundaries and finite
  geometry induce a remarkably rich phase behavior of the confined fluid, exhibiting three distinct types of condensation, reentrant phenomena, and
  continuous higher-order transitions between condensation states. Through a detailed macroscopic analysis, we derive the conditions for each type of
  condensation, classify the global phase diagrams, and explore asymptotic behavior in specific limiting cases. The asymptotic predictions are
  confirmed by a detailed comparison with the  numerical solutions of the governing equations.
\end{abstract}

\maketitle

\section{Introduction}

Wetting phenomena and capillary condensation are two prominent examples of phase transitions in fluids induced by the presence of a spectator phase,
such as a solid wall or a pair of walls forming a slit. These phenomena illustrate the interplay of surface tension with bulk fluid properties, the
competition between wall-fluid and fluid-fluid microscopic interactions, and the subtle roles of interfacial fluctuations and finite-size scaling. A
detailed understanding of their properties has been the subject of intense scrutiny over several decades, described in numerous monographs and
reviews (see, for example, \cite{RW, gregg, croxton, sull, dietrich, evans90, charvolin, schick, forgacs, henderson92, israel, gelb, bonn}).

While the simple planar models provide crucial insights into these phenomena, their extensions to systems with a broken translation symmetry induce
novel types of phase behavior. A notable example is a confinement formed by two \emph{unparallel} plates making an angle $\alpha$ , forming a linear
wedge \cite{hauge, rejmer, wood, abraham, bruschi, our_wedge, delfino, our_wedge2}. If the bulk pressure $p$ (or equivalently the chemical potential
$\mu$) is increased towards its saturation value $p_{\rm sat}$ ($\mu_{\rm sat}$) at a fixed subcritical temperature $T$, the fluid inside the wedge
condenses continuously, akin to complete wetting on a single wall. There are two major differences though. First, while complete wetting requires
$T\ge T_w$, where the wetting temperature $T_w$ is associated with the vanishing Young's contact angle $\theta$ of a sessile macroscopic droplet,
complete filling in a linear wedge occurs at a lower temperature $T_f<T_w$. This filling temperature is determined by both the contact angle and the
opening angle, such that \cite{hauge}
\begin{equation}
\theta(T_f)=\frac{\pi-\alpha}{2}\,. \label{filling}
\end{equation}
Second, unlike complete wetting, where critical singularities (in three dimensions) are governed by microscopic forces \cite{lipowsky84, lipowsky87},
complete filling is dominated by the system geometry. Specifically, the divergence of adsorption in wedges can be largely described by the motion of
a meniscus separating the condensed phase from the outer gas, which retreats from the apex as the system approaches saturation \cite{hauge, rejmer,
  wood, our_wedge}.

\begin{figure*}[bht]
  \subfloatflex{Single pinning (SP)}{%
    \includegraphics[scale=0.5]{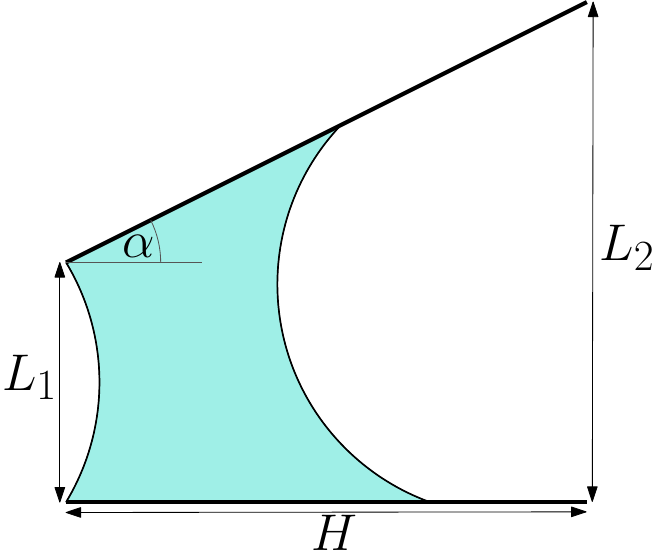}%
  } \hspace*{0.2cm}
  \subfloatflex{Semi-double pinning (SDP)}{%
    \includegraphics[scale=0.5]{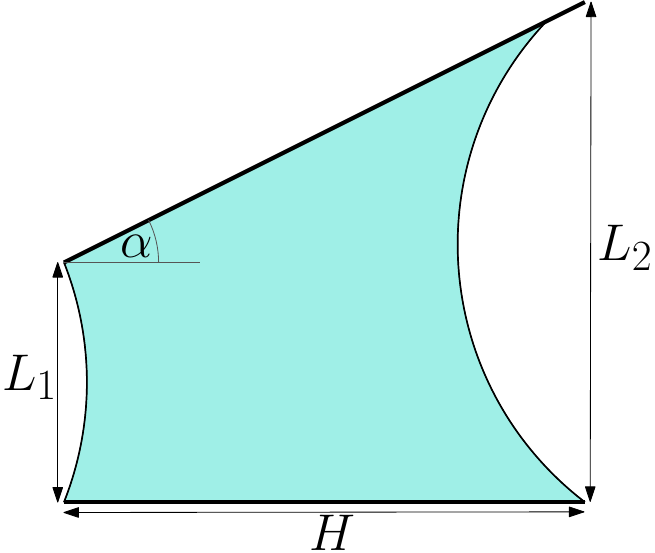}%
  }\hspace*{0.2cm}
  \subfloatflex{Double pinning (DP)}{%
    \includegraphics[scale=0.5]{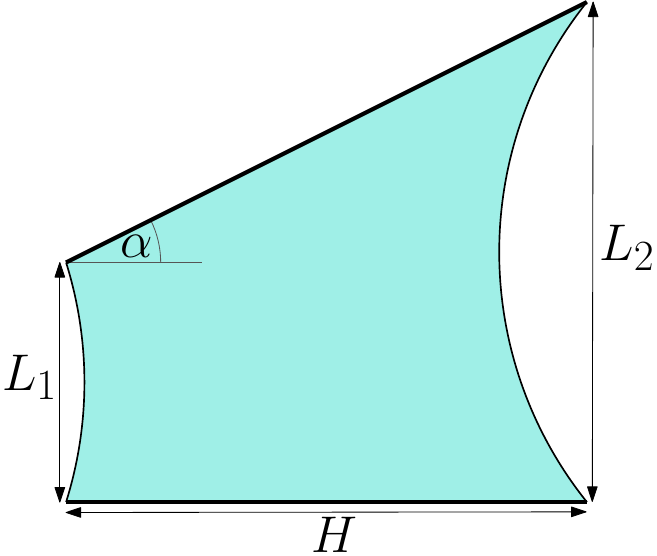}%
  }
  \caption{Schematic representation of the three possible condensation states in open wedges, depicting the distinct meniscus configurations characteristic of each state.
  The illustrations correspond to two-dimensional cross-sections of macroscopically deep wedges, where translation invariance along the direction normal to the plane is assumed.} \label{sketch}
\end{figure*}

However, an even greater contrast arises when comparing condensation in wedges to that in planar capillary slits, where the latter exhibits a
first-order transition. Capillary condensation refers to an abrupt transition from a capillary gas to a capillary liquid, which, according to the
classical Kelvin equation, occurs at the chemical potential $\mu_{\rm cc}=\mu_{\rm
    sat}-\delta\mu_{\rm cc}$, where
\begin{equation}
\delta\mu_{\rm cc}(L)=\frac{2\gamma\cos\theta}{L\Delta\rho}  \label{kelvin_slit}
\end{equation}
and terminates at a capillary critical temperature $T_c(L)$ \cite{nakanishi81, nakanishi83}. Here, $\Delta\rho=\rho_l-\rho_g$ is the difference in
bulk liquid and gas densities, $L$ is the slit width  and $\theta$ is Young's contact angle, assumed to be less than $\pi/2$. The Kelvin equation
arises directly from a geometric condition, which states that at the chemical potential $\mu_{\rm cc}$ a capillary gas and a capillary liquid can
coexist, separated by a curved meniscus with a Laplace radius of curvature $R=\gamma/(\delta\mu_{\rm cc}\Delta\rho)$ connecting the slit walls at
Young's contact angle $\theta$.

Thus far, we have been assuming that the confining walls are infinitely long. Relaxing this assumption by considering finite wall extensions
introduces significant changes to the phase behavior. For instance, complete wetting is precluded on a finite-width wetting stripe, despite still
exhibiting some scaling behavior \cite{chmiel, lenz, gau, bauer, frink, hend_99, bauer2, lipowsky_01, dietrich5, porcheron, berim, our_stripe,
  our_stripes, pos23}. If two walls of a finite length $H$ are placed face-to-face at a distance $L$, capillary condensation still persists, provided
$H>L\sec\theta$, but the chemical potential shift is now reduced and described by a modified Kelvin equation \cite{our_slit}
\begin{equation}
\delta\mu_{\rm cc}(L,H)=\frac{2\gamma\cos\theta_e}{L\Delta\rho}\,.  \label{kelvin_finite}
\end{equation}
where $\theta_e$, the so-called edge contact angle, is determined by the system's aspect ratio $H/L$
\begin{equation}
\cos\theta_e=\cos\theta-\frac{L}{2H}\left[\sin\theta_e+\sec\theta_e\left(\frac{\pi}{2}-\theta_e\right)\right]\,. \label{theta_e}
\end{equation}
The value of $\theta_e$ given  Eq.~(\ref{theta_e}) describes the contact angle of the menisci formed at the slit openings at the point of the phase
transition.

The purpose of this work is to examine the phase behavior of fluids confined in an \emph{open} linear wedge. This model can be envisioned as a finite
section of a linear wedge of an opening angle $\alpha$, bounded by two slices of horizontal distance $H$. The widths of the openings, hereafter
referred to as left and right, are denoted $L_1$ and $L_2$, respectively, are related by $L_2=L_1+H\tan\alpha$.  The system retains macroscopic
depth, ensuring translational invariance along the direction normal to the wedge cross-section.  This model can also be viewed as a modification of a
slit formed by nonparallel walls with axial symmetry \cite{jj}, and it will be shown that the breaking of this symmetry gives rise to even more
intricate phase behavior.

Our objective is to demonstrate that the phase behavior in open wedges fundamentally differs from that of infinitely long wedges. Most importantly,
we show that, unlike wedge filling, condensation in open wedges is always associated with a first-order jump. Depending on the system
parameters---namely $\alpha$, $\theta$ and the aspect ratio $a=H/L_1$---we identify three distinct types of condensation. The macroscopic
condensation state can be described by the shape and location of two menisci separating the condensed phase from the bulk gas. While the left
meniscus is always pinned at the narrow opening, the location of the right meniscus determines the condensation state:

\begin{enumerate}

  \item For systems with small aspect ratios, the right meniscus is pinned at the wider opening,
        resulting in condensation throughout the entire volume analogous to capillary condensation in finite slits (see Fig.~1c). This state will henceforth
        be referred to as \emph{double pinning} (DP).

  \item In an alternative scenario, first-order condensation can result in a partially filled system, where the right meniscus resides within
        the wedge and meets the walls at the equilibrium Young's contact angle (see Fig.~1a). We shall hereafter refer to this state as \emph{single pinning}
        (SP). As the pressure increases, the meniscus gradually advances toward the wider opening, in some analogy with the process of complete filling.

  \item Finally, an intermediate state exists where the right meniscus is pinned at one wall edge while remaining free
        to slide along the other wall (see Fig.~1b). We will refer to this state as \emph{semi-double pinning} (SDP).

\end{enumerate}

The remainder of this paper is organized as follows: In Sec. II, we present a macroscopic description of the three condensation types, determining
the corresponding grand potentials, phase boundaries, and the nature of the phase transitions. Section III introduces necessary conditions to
constrain possible condensation scenarios. In Sec. IV, we construct global phase diagrams and categorize them into four qualitatively distinct
classes. Section V discusses asymptotic properties in specific limiting cases. We conclude in Sec. VI with a summary of our findings and highlight
possible extensions of this work.

\section{Three Condensation Regimes} \label{regimes}

We begin by formulating the macroscopic conditions for the three admissible condensation regimes, as illustrated in Fig.~1. While these regimes
differ in the location of the right meniscus, the left meniscus is always pinned at the narrow (left) opening of width $L_1$. The shapes of the
menisci are characterized by their Laplace radii of curvature, $R$, and the contact angles at which they meet the walls. Specifically, the left
meniscus forms contact angles $\theta_1$ (with the bottom wall) and $\theta_1'$ (with the top wall), which are tied by the relation
$\theta'_1=\theta_1+\alpha$ evident from Fig.~2.

Additionally, the system geometry dictates that
\begin{equation}
\cos\theta_1=\frac{L_1}{2R}\,, \label{cos_theta1}
\end{equation}
which establishes the pressure dependence of $\theta_1$. Although Eq.~(\ref{cos_theta1}) applies to all condensation states, we are in particular
interested in such a value of $\theta_1$ which corresponds to a condensation to the given regime. This value is determined by the free-energy
balance, $\Omega^{\rm ex}=0$, where $\Omega^{\rm ex}$ is the excess grand potential per unit length relative to the low-density (gas-like) state.
Macroscopically, $\Omega^{\rm ex}$ can be expressed as
\begin{equation}
\Omega^{\rm ex}=\delta pS+\gamma(\ell_{\rm left}+\ell_{\rm right})-\ell_w\gamma\cos\theta\,. \label{om_ex}
\end{equation}
Here, the first term represents the free-energy cost associated with the presence of a metastable liquid, with pressure $p_l=p-\delta p$, occupying
the cross-sectional area $S$. The second term accounts for the interfacial energy due to the formation of the left meniscus of length $\ell_{\rm
left}$
\begin{equation}
\ell_{\rm left}=R(\pi-2\theta_1)\,, \label{ell_left}
\end{equation}
and the right meniscus of length $\ell_{\rm right}$, the specific form of which depends on the condensation regime and is described below. The final
term in Eq.~(\ref{om_ex}) represents the difference in surface energies between the wall-liquid and wall-gas interfaces, with a total length
$\ell_w$, where Young's law has been applied.

First-order condensation transitions occur at the chemical potential $\mu_i$, where the subscript specifies the condensation regime: SP, SDP, or DP.
Expanding $\delta p\approx\delta\mu\Delta\rho$, where $\delta\mu_i=\mu_{\rm sat}-\delta\mu_i$ specifies the departure from saturation, a Kelvin-like
equation that  describes all three condensation regimes is of the same form
\begin{equation}
\delta\mu_i=\frac{2\gamma\cos\theta_1}{\Delta\rho L_1}\,,\;\;\;(i={\rm SP\,,\;SDP\,,\;DP})\,. \label{kelvin_new}
\end{equation}
To this end, the value of $\theta_1$  is sufficient to describe the system phase behaviour for which, however, we need to specify the configuration
of the right meniscus. In the following, we outline the procedure for each condensation regime separately.

\subsection{Single pinning regime} \label{sub_sp}

\begin{figure}[t]
  \centering
  \includegraphics[scale=0.39]{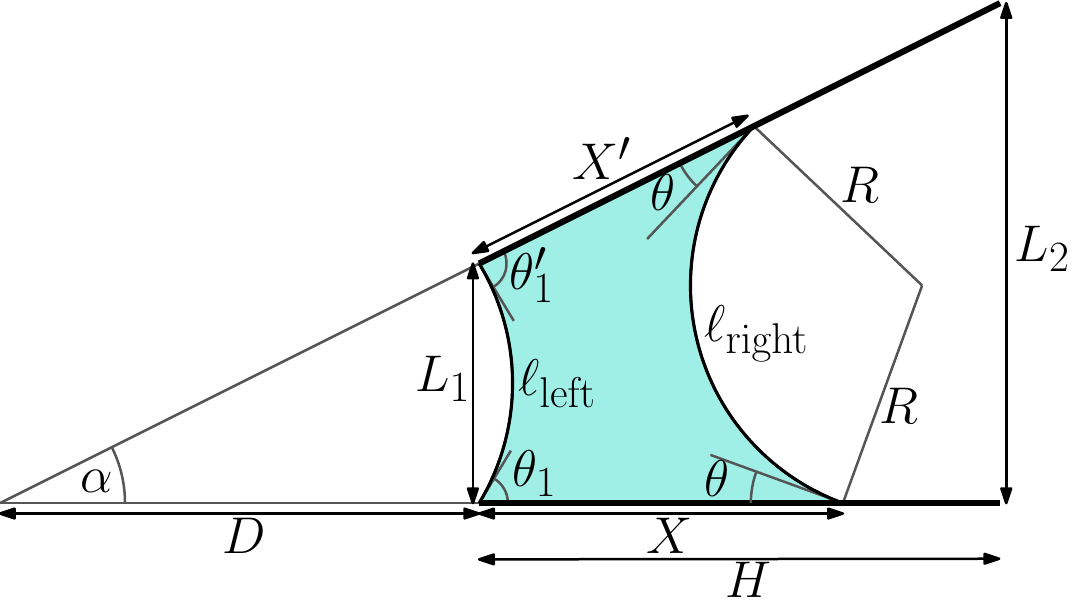}
  \caption{Schematic representation of the single pinned (SP) condensation state, illustrating the key lengths and angles used in the derivation
    of $\Omega^{\rm ex}_{\rm SP}$. Specifically, $\ell_{\rm
        left}$ and $\ell_{\rm right}$ denote the arc lengths of the left and right menisci, respectively, both with the Laplace radius $R$. Additionally, $X$ and $X'$ represent the lengths of the walls in
    contact with the liquid. } \label{fig_single}
\end{figure}

In sufficiently large systems, condensation can occur at low pressures (or chemical potentials), leading to the formation of highly curved menisci
and enabling the SP condensation regime (see Fig.~\ref{fig_single}). This regime is characterized by the depinned right meniscus, which meets both
walls at the equilibrium contact angle $\theta$. The position of the right meniscus is defined by the distance $X$ between its contact point on the
bottom wall and the left corner, which follows from the relationship
\begin{equation}
\tan\alpha=\frac{R[\cos\theta+\cos(\theta+\alpha)+\sin(\alpha+\theta)\tan\alpha]}{D+X+R\sin\theta}\,,
\end{equation}
where $D=L_1\cot\alpha$ (see Fig.~\ref{fig_single}), implying
\begin{multline}
  X=R\cot\alpha[\cos\theta+\cos(\theta+\alpha)]\\+R\sin(\alpha+\theta)-R\sin\theta-L_1\cot\alpha\,. \label{X}
\end{multline}
From here and using the relation
\begin{equation}
\cos\alpha=\frac{X+R\sin\theta}{X'+R\sin(\alpha+\theta)\sec\alpha}\,,
\end{equation}
we can also determine the contact length between liquid and the upper wall
\begin{equation}
X'=\csc\alpha[R\cos\theta+R\cos(\theta+\alpha)-L_1]\,. \label{XX}
\end{equation}
Thus, for the total contact length $\ell_w=X+X'$ we obtain
\begin{multline}
  \ell_w=R[\sin(\alpha+\theta)-\sin\theta]\\
  +\frac{1+\cos\alpha}{\sin\alpha}[R\cos\theta+R\cos(\theta+\alpha)-L_1]\,. \label{ellw_sp}
\end{multline}
Furthermore, the area $S$ of the condensed region can be expressed as
\begin{multline}
  \hspace{-\multlinegap}S=R^2\left[\sin\theta_1\cos\theta_1-\left(\pi-\theta_1-\theta-\frac{\alpha}{2}\right)+\frac{1}{2}\sin\theta\cos\theta\right]\\
  -\frac{R^2}{2}\left[\sin(\alpha+\theta)\cos(\alpha+\theta) +\sin^2(\alpha+\theta)\tan\alpha\right]\\
  +\frac{X+R\sin\theta}{2}\left[L_1 + R\cos\theta+R\cos(\alpha+\theta)\right]\\
  +\frac{X+R\sin\theta}{2}R\sin(\alpha+\theta)\tan\alpha\label{S_sp}\,.
\end{multline}
Finally, the right meniscus arc length is given by
\begin{equation}
\ell_{\rm right}=R(\pi-2\theta-\alpha)\,. \label{ell_right_sp}
\end{equation}

Substituting from Eqs.~(\ref{ell_left}), (\ref{ellw_sp}), (\ref{S_sp}), and (\ref{ell_right_sp}) into Eq.~(\ref{om_ex}), the excess grand potential for an SP state is:
\begin{multline}
  \hspace{-\multlinegap}\Omega^{\rm ex}_{\rm SP}=   R\left(\cos\theta \sin\theta+\cos\theta_1 \sin\theta_1\right)\\
  -\frac{R \cos\alpha \cos^2\theta}{\sin\alpha} - \frac{R \cos^2\theta}{\sin\alpha}
  + R\left(\pi  - \frac{1}{2} \alpha -  \theta -  \theta_1\right)\\
  +\frac{L_1 \cos\alpha \cos\theta}{\sin\alpha} - \frac{L_{1}^{2} \cos\alpha}{2 R \sin\alpha} + \frac{L_1 \cos\theta}{\sin\alpha}\,.    \label{omex_sp}
\end{multline}
Setting $\Omega_{\rm SP}^{\rm ex}=0$, we get the criterion for SP condensation:
\begin{multline}
  \hspace{-\multlinegap}\frac{1}{2}\left(\cot\alpha +\csc \alpha \right) \cos^{2}\theta -(\cot\alpha+\csc\alpha)\cos\theta_1\cos\theta
  \\-\frac{1}{2}\left(\sin\theta_1\cos\theta_1+\sin\theta\cos\theta\right) +\cos^2\theta_1\cot\alpha\\
  -\frac{\pi}{2}+\frac{\alpha}{4}+\frac{\theta}{2}+\frac{\theta_1}{2}=0\,, \label{kelvin_single2}
\end{multline}
where Eq.~(\ref{cos_theta1}) has been used. This implicit equation for $\theta_1$ determines the location of SP condensation through Eq.~(\ref{kelvin_new}).

\subsection{Semi-double pinning regime}

\begin{figure}[t]
  \centering
  \includegraphics[scale=0.39]{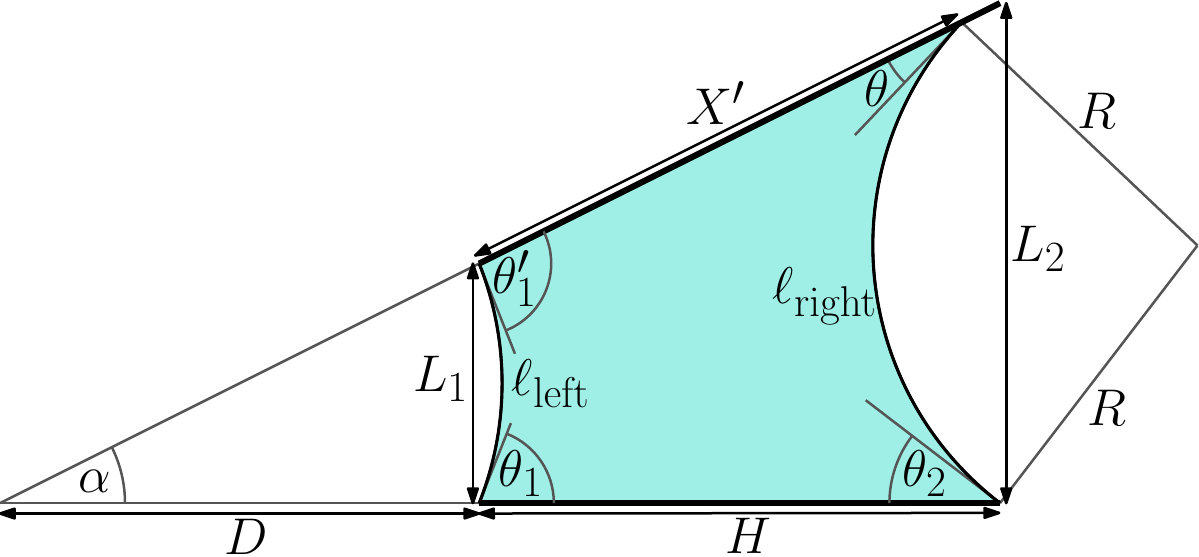}
  \caption{Schematic representation of the semi-double pinned (SDP) condensation state, illustrating the key lengths and angles used in the derivation
    of $\Omega^{\rm ex}_{\rm SDP}$. Specifically, $\ell_{\rm left}$ and $\ell_{\rm right}$ denote the arc lengths of the left and right menisci,
    respectively, both with the Laplace radius $R$. Additionally, $X'$ represents the length of the upper wall in contact with the
    liquid.}\label{fig_semi_double}
\end{figure}

In the SDP condensation regime, the right meniscus remains pinned at the right bottom edge but connects the upper wall inside the wedge at the
equilibrium contact angle $\theta$ (see Fig.~\ref{fig_semi_double}). This introduces two independent edge contact angles: $\theta_1$, which
characterizes the shape of the left meniscus and is determined by Eq.~(\ref{cos_theta1}), and $\theta_2$, the angle that the right meniscus forms
with the bottom wall, satisfying (see Appendix \ref{app_sdp}):
\begin{equation}
\frac{L_2}{R}=\frac{\cos\theta+\cos(\theta_2+\alpha)}{\cos\alpha}\,. \label{sd_theta2}
\end{equation}
In this condensation regime, the key measures in Eq.~(\ref{om_ex}) are given by
 \begin{multline}
\ell_w=H(1+\sec\alpha)\\-\csc\alpha[L_2-R(\cos\theta_2+\cos(\theta+\alpha))]\,,
\end{multline}
\begin{equation}
\ell_{\rm right}=R(\pi-\theta-\theta_2-\alpha)
\end{equation}
and
\begin{multline}
  \hspace{-\multlinegap}S=R^2\left[\sin\theta_1\cos\theta_1-\left(\pi-\theta_1-\frac{\theta}{2}-\frac{\theta_2}{2}-\frac{\alpha}{2}\right)\right]\\
  -\frac{R^2}{2}\left[\sin(\alpha+\theta)\cos(\alpha+\theta)+\sin^2(\alpha+\theta)\tan\alpha\right]\\
  -\frac{R^2}{2}\sin\theta_2\cos\theta_2+\frac{H+R\sin\theta_2}{2}\left(L_1+R\cos\theta_2\right)\\
  +\frac{H+R\sin\theta_2}{2}\left[R\cos(\alpha+\theta)+R\sin(\alpha+\theta)\tan\alpha\right]\,,
\end{multline}
in addition to  Eq.~(\ref{ell_left}) for $\ell_{\rm left}$.

Substituting these into Eq.~(\ref{om_ex}), we get for the excess grand potential in the SDP state
\begin{multline}
  \hspace{-\multlinegap}\Omega^{\rm ex}_{\rm SDP}=  \frac{R\cos\theta \sin\theta}{2} -\frac{R \cos\alpha \cos^2\theta}{2 \sin\alpha} + R \cos\theta_1 \sin\theta_1\\
  + R\left(\pi  - \frac{\alpha}{2} - \frac{\theta}{2} - \theta_{1} - \frac{\theta_{2}}{2}\right) - H \cos\theta + \frac{H \cos\theta_2}{2} \\
  + \frac{L_1 \sin\theta_2}{2}- \frac{R \cos\theta \cos\theta_2}{\sin\alpha} + \frac{R \cos\theta \sin\theta_2}{2 \cos\alpha} + \frac{H L_1}{2 R} \\
  - \frac{H \cos\theta}{2 \cos\alpha}
  + \frac{L_2 \cos\theta}{\sin\alpha} - \frac{R \cos\theta^{2}}{2 \cos\alpha \sin\alpha}\,.  \label{omex_sdp}
\end{multline}

The SDP condensation occurs when $\Omega^{\rm ex}_{\rm SDP}=0$, leading to
\begin{multline}
  \hspace{-\multlinegap}\frac{\left[2 \csc  \alpha\left(L_2-R \cos \theta_2 \right)+\left(R \sin \theta_2-H \right) \sec\alpha -2 H \right] \cos
    \theta}{R}
  \\+\sin\theta\cos\theta-\left(\sec \alpha  \csc \alpha+\cot \alpha \right) \cos^2\theta +2\pi-\alpha-\theta-2\theta_1-\theta_2\\
  +2\left(\sin \theta_1   + \sin \theta_2 +\frac{H}{R} \right) \cos \theta_1 +\frac{H \cos \theta_2 }{R}=0\,.
  \label{sd_therm}
\end{multline}

As a result,  we have a system of three equations---(\ref{cos_theta1}), (\ref{sd_theta2}), and (\ref{sd_therm})---for the three unknowns $\theta_1$,
$\theta_2$, and $R$, determining the location of SDP condensation. Alternatively, we may eliminate the length variables, to obtain only a set of two
equations:
\begin{multline}
  \hspace{-\multlinegap}\left(\cot \alpha+2 \csc \alpha\right) \cos^{2}\theta-\sin\theta\cos\theta-\sin(2\theta_1) +\cos \theta_2 \sin \theta_2\\
  +\left[2\cot\alpha\left(\cos \theta_2-2\cos  \theta_1\right)-4 \cos  \theta_1 \csc \alpha-2
    \sin \theta_2\right] \cos
  \theta\\
  +\left(4 \cos^{2}\theta_1-\cos^{2}\theta_2\right) \cot \alpha=2 \pi -\alpha -\theta -2 \theta_1
  -\theta_2
  \label{sd_theta2b}
\end{multline}
and
\begin{equation}
\cos\theta_1=\frac{L_1}{2L_2}\frac{\cos(\theta_2+\alpha)+\cos\theta}{\cos\alpha}\,, \label{sd_theta1b}
\end{equation}
involving only the contact angles.

\subsection{Double pinning regime}

\begin{figure}[t]
  \includegraphics[scale=0.39]{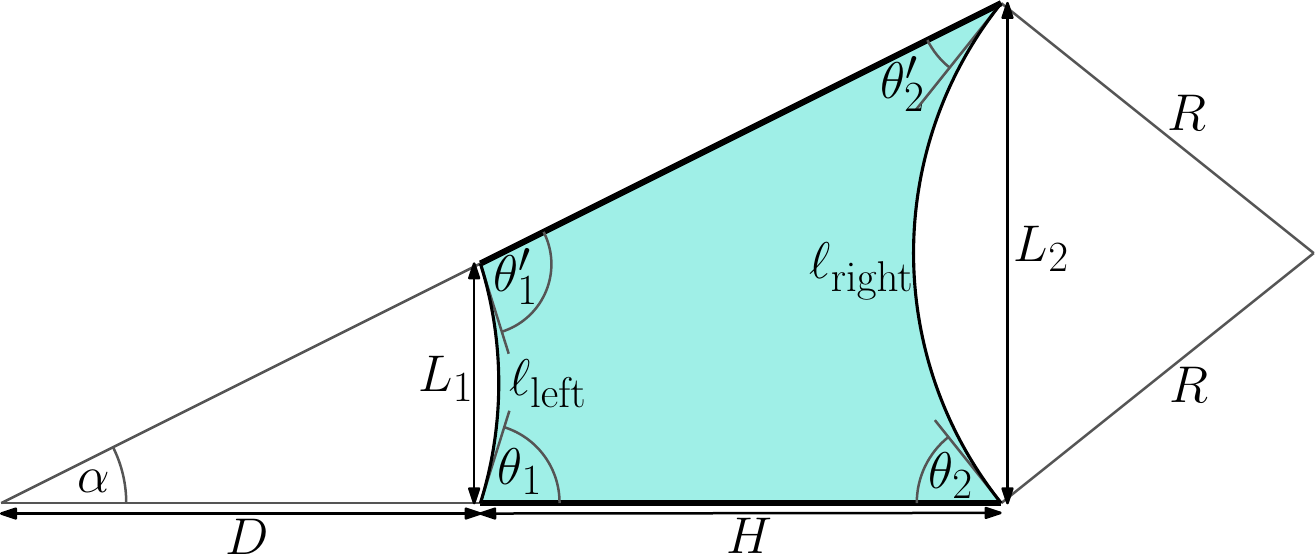}
  \caption{Schematic representation of the double pinned (DP) condensation state, highlighting the key lengths and angles used in the derivation of
    $\Omega^{\rm ex}_{\rm DP}$. Specifically, $\ell_{\rm left}$ and $\ell_{\rm right}$ represent the arc lengths of the left and right menisci,
    respectively, both with the Laplace radius $R$.} \label{fig_double}
\end{figure}

Finally, we turn to the DP regime, where both menisci are pinned at their respective edges. A description of this state requires again a pair of
contact angles, $\theta_1$ and $\theta_2$ where for the latter we have the relation
\begin{equation}
\cos\theta_2=\frac{L_2}{2R}\,, \label{dp_theta2}
\end{equation}
analogous to Eq.~(\ref{cos_theta1}) for $\theta_1$.  Again, $\theta_2$ represents the contact angle at the bottom wall, and it is related to the
contact angle at the top wall by $\theta_2'=\theta_2-\alpha$, as illustrated in Fig.~\ref{fig_double}.

In this condensation regime, the total arc length of the menisci can be expressed as
\begin{equation}
\ell_{\rm left}+\ell_{\rm right}=2R(\pi-\theta_1-\theta_2)\,,
\end{equation}
the area occupied by the liquid  as
\begin{eqnarray}
  S&=&[L_1+L_2]H/2-R^2(\pi-2\theta_1)/2+R^2\sin\theta_1\cos\theta_1\nonumber\\
  &&-R^2(\pi-2\theta_2)/2+R^2\sin\theta_2\cos\theta_2\,,
\end{eqnarray}
while length of the wall-liquid contact is simply
\begin{equation}
\ell_w=H(1+\sec\alpha)\,.
\end{equation}

Substituting these expressions into Eq.~(\ref{om_ex}) yields the excess grand potential for the DP state
\begin{multline}
  \Omega^{\rm ex}_{\rm DP}= \frac{(L_1+L_2)H}{2R}+R(\sin\theta_1\cos\theta_1+\sin\theta_2\cos\theta_2)\\
  +R(\pi-\theta_1-\theta_2)-H(1+\sec\alpha)\cos\theta\,,  \label{omex_dp}
\end{multline}
implying  the following condition for the DP condensation
\begin{multline}
  (\cos\theta_1+\cos\theta_2)H+R(\pi-\theta_1-\theta_2)-H(1+\sec\alpha)\cos\theta\\
  +R\sin(\theta_1+\theta_2)\cos(\theta_1-\theta_2)=0\,. \label{cc_dp}
\end{multline}

The set of equations (\ref{cos_theta1}), (\ref{dp_theta2}), and (\ref{cc_dp}) determine the location of the DP condensation.

\subsection{Phase boundaries between the condensed states}

In the final part of this section, we discuss the locations and properties of phase transitions between the three condensed regimes. First, we claim
that the transitions must necessarily be continuous, for the following reason. Consider the case of completely wet walls ($\theta = 0$), where the
right meniscus connects tangentially to the walls and moves smoothly along them as the pressure increases toward saturation, precluding any
possibility of a jump in the meniscus position. In this case, the system evolves continuously from the SP state to the SDP state and then to the DP
state, provided all states are accessible (discussed later). However, this also implies that the transitions must be continuous for $\theta> 0$,
since any discontinuity in the meniscus position would be even less favorable than in the case of complete wetting.

It is straightforward to determine the phase boundaries separating the successive regimes. Let us first assume that the system is in the SP condensed
regime. As the pressure increases, the right meniscus is pushed away from the left one, which remains pinned at the left opening, and slides toward
the other end of the wedge. This process terminates when the right meniscus becomes pinned at the lower right edge, signaling the transition to the
SDP regime. Thus, the SP/SDP boundary is given by the condition $\theta_2 = \theta$, or equivalently, $X = H$. According to Eq.~(\ref{sd_theta2}),
this occurs at the  pressure $p_{\rm SP-SDP} = p_{\rm sat} - \delta p_{\rm SP-SDP}$, where
\begin{equation}
\delta p_{\rm SP-SDP}=\frac{\gamma[\cos\theta+\cos(\alpha+\theta)]}{L_2\cos\alpha}\,. \label{p_sp_sdp}
\end{equation}
At the transition the grand potentials of the two regimes are equal, i.e., $\Delta_{\mathrm{SP-SDP}}=\Omega^{\rm ex}_{\rm SP}-\Omega^{\rm ex}_{\rm
SDP}=0$ and the nature of the transition further implies that the first derivative of $\Delta_{\mathrm{SP-SDP}}$ with respect to $R$ also vanishes at
$R = R_{\rm SP-SDP} = \gamma / \delta p_{\rm SP-SDP}$,  while for the second derivative we obtain
\begin{equation}
\pdiff[2]{\left(\Delta_{\mathrm{SP-SDP}}\right)}{R}\Big|_{R_{\rm SP-SDP}}=\frac{2{\left[\cos\!\left(\alpha + \theta\right) + \cos\!\theta\right]}^{3}
  \sin\!\theta}{L_{2} \sin\!\left(2\alpha\right) \sin\!\left(\alpha + \theta\right)}\,,
\end{equation}
which is zero for $\theta=0$. The first non-zero derivative for $\theta=0$ is the third derivative, which at the  transition point is
\begin{equation}
  \pdiff[3]{\left(\Delta_{\mathrm{SP-SDP}}\right)}{R}\Big|_{R_{\rm SP-SDP}}=\frac{4\cos^{7}\!\left(\frac{\alpha}{2}\right)}{L_{2}^{2} \cos^{2}\!\alpha
    \sin^{3}\!\left(\frac{\alpha}{2}\right)}\,,\;\;\;(\theta=0)\,.
\end{equation}

As the pressure increases further, the right meniscus remains pinned at the lower edge but is free to move along the upper wall. This regime
terminates when the right meniscus gets pinned at the upper right edge, which occurs when $\theta_2' = \theta$ (and $X' = H\sec \alpha$). Since
$\theta_2' = \theta_2 - \alpha$, it follows from Eq.~(\ref{dp_theta2}) that this second pinning occurs at the pressure $p_{\rm SDP-DP} = p_{\rm sat}
  - \delta p_{\rm SDP-DP}$, where
\begin{equation}
\delta p_{\rm SDP-DP}=\frac{2\gamma\cos(\alpha+\theta)}{L_2}\,. \label{p_sdp_dp}
\end{equation}
Comparing with (\ref{p_sp_sdp}) it can be verified that $p_{\rm SDP-DP}>p_{\rm SP-SDP}$, as required.  At  the  SDP/DP transition point  both
$\Delta_{\mathrm{SDP-DP}}=\Omega^{\rm ex}_{\rm SDP}-\Omega^{\rm ex}_{\rm DP}$ and its first derivative with respect to $R$ vanish, while the second
derivative is
\begin{equation}
  \pdiff[2]{\left(\Delta_{\mathrm{SDP-DP}}\right)}{R}\Big|_{R_{\rm SDP-DP}}=\frac{4\cos^{4}\!\left(\alpha + \theta\right) \sin\!\theta}{L_{2}
    \sin\!\left(2\alpha + \theta\right) \sin\!\left(\alpha + \theta\right)}\,,
\end{equation}
which again vanishes for $\theta = 0$. For completely wet walls, the next derivative of the grand potential difference is
\begin{equation}
  \pdiff[3]{\left(\Delta_{\mathrm{SDP-DP}}\right)}{R}\Big|_{R_{\rm SDP-DP}}=\frac{4\cos^{5}\!\alpha}{L_{2}^{2} \sin^{3}\!\alpha}\,,\;\;\;(\theta=0)\,,
\end{equation}
which is finite even for $\theta = 0$.

We conclude this section with a brief summary. The system exhibits three condensed regimes: SP, SDP, and DP, which are defined by the location of the
right meniscus. At pressures far from saturation, when $\delta p>\delta p_{\rm SP-SDP}$, with $\delta p_{\rm SP-SDP}$ given by Eq.~(\ref{p_sdp_dp}),
the system may condense into the SP regime if the condition (\ref{kelvin_single2}) is satisfied. As the pressure increases further, the system
undergoes the SP-SDP phase transition at the pressure $p_{\rm SP-SDP}$, followed by the SDP-DP transition at the pressure $p_{\rm SDP-DP}$. Both of
these transitions, characterized by the pinning of the right meniscus to the lower and upper right edges, respectively, are continuous. According to
the classical Ehrenfest classification, they are second-order transitions if the confining walls are partially wet, and third-order transitions if
the walls are completely wet.

However, this is not the only possible scenario. For instance, the condition (\ref{kelvin_single2}) may not be satisfied for any pressure $p<p_{\rm
SP-SDP}$, and the system may condense directly into the SDP regime for $p_{\rm SP-SDP}< p< p_{\rm SDP-DP}$, as described by the free-energy balance
in Eq.~(\ref{sd_theta2b}). In the SDP regime, the system may still undergo the continuous depinning transition at the pressure $p_{\rm SDP-DP}$ to
reach the DP state, or it may remain in the SDP state even at saturation.  The latter occurs when $p_{\rm SDP-DP} \geq 0$. Another possibility is
that the condensation bypasses both the SP and SDP regimes and directly adopts a DP state at a pressure $p> p_{\rm SDP-DP}$. Beyond this point, no
further phase transition occurs upon further increase of pressure and the menisci just gradually flatten. Finally, it is also possible that the
system does not condense at all before bulk condensation. In the following two sections, we will explore all possible scenarios which can realize
depending on the system parameters $\alpha$, $\theta$, and $a=H/L_1$.

\section{Limits of condensation}

Before discussing the complete phase behavior of the system, let us examine the scenario at the highest possible pressure prior to bulk condensation,
$p=p_{\rm sat}^-$. At this pressure limit, the system can exist in only three possible states:
\begin{enumerate}
  \item[i)] the DP  state,
  \item[ii)] the SDP  state, or
  \item[iii)] a non-condensed state.
\end{enumerate}

To show this, we go through all possibilities in detail below.

\subsection{DP regime} \label{dp_limit}

\begin{figure}[t]
  \includegraphics[scale=0.5]{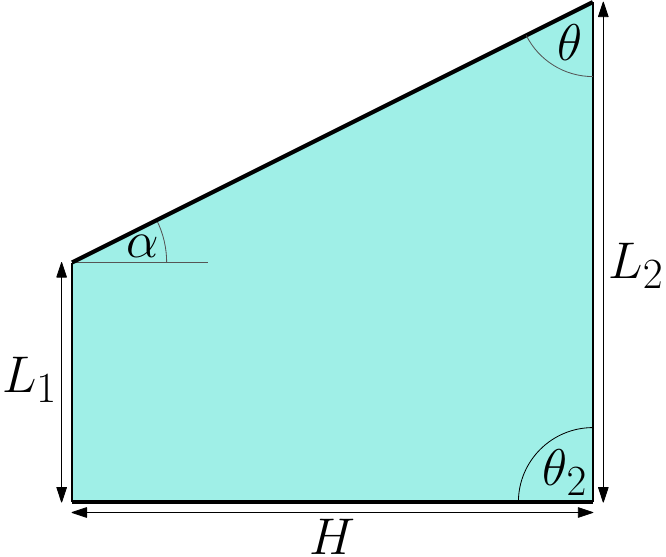}
  \caption{Schematic representation of the DP state at saturation, where the meniscus is on the verge of depinning from the upper wall. }
  \label{sdp_rinfty}
\end{figure}

In the DP state, recall, both menisci are pinned, with edge contact angles  $\theta_1$ and $\theta_2$ determined  by Eqs.~(\ref{cos_theta1}) and
(\ref{dp_theta2}), respectively. In the limit $\delta p\to0$, where the curvature radius $R\to\infty$, the edge contact angles approach right angles
as
\begin{eqnarray}
  &\theta_1&\to\frac{\pi}{2}-\frac{L_1}{2R}\,, \;\;\; {\rm as}\; R\to\infty \label{theta1_lim}\\
  &\theta_2&\to\frac{\pi}{2}-\frac{L_2}{2R}\,, \;\;\; {\rm as}\; R\to\infty\,.  \label{theta2_lim}
\end{eqnarray}
Substituting these into Eq.~(\ref{cc_dp}), we obtain
\begin{equation}
2L_1+H[\tan\alpha -(1+\sec\alpha)\cos\theta]+{\cal{O}}(R^{-1})=0\,, \label{asym_dp}
\end{equation}
The leading zeroth-order contribution  corresponds to flat menisci (when $\delta p=0$) and establishes the minimum system length $H^{\rm DP}_{\rm
      min}$, relative to $L_1$, required for DP condensation
\begin{equation}
a^{\rm DP}(\theta,\alpha)=\frac{H^{\rm DP}_{\rm min}}{L_1}=\frac{2}{(1+\sec\alpha)\cos\theta-\tan\alpha}\,. \label{amin_dp}
\end{equation}
Note that in the limit $\alpha\to0$, this simplifies to $H=L_1\sec\theta$, recovering the condition for the minimum length-to-width ratio allowing
condensation in finite slits.

From Eq.~(\ref{amin_dp}), further restrictions on DP condensation can be derived purely in terms of the geometric and contact angles. Specifically,
the positivity of $a^{\rm DP}$ implies
\begin{equation}
\cos\theta>\frac{\sin\alpha}{1+\cos\alpha}\,, \label{dp_angle}
\end{equation}
indicating that DP condensation is favored for small opening angles $\alpha$ or for small contact angles $\theta$ when the geometry is fixed.

However, a more stringent constraint on the DP condensation regime is obtained by comparing the stability of the DP and SDP regimes at saturation.
The sketch in Fig.~\ref{sdp_rinfty} illustrates the marginal DP configuration when $\theta_2 = \pi/2$ (required by the condition $\delta p=0$) and
simultaneously $\theta_2=\pi-\alpha-\theta$ (indicating the depinning of the meniscus at the upper wall). Therefore, the necessary condition for DP
condensation is
\begin{equation}
\alpha+\theta<\frac{\pi}{2}\;\;\;{\rm (DP)}\,,
\end{equation}
or, equivalently,
\begin{equation}
\sin\alpha<\cos\theta\;\;\;{\rm (DP)}\,,
\end{equation}
which clearly provides a stricter condition than Eq.~(\ref{dp_angle}).

\subsection{SDP regime} \label{sdp_limit}

In the SDP regime,  the free-energy balance in the limit of $\delta p=0$ simplifies to
\begin{equation}
L_1+\ell_{\rm right}=(H+X')\cos\theta\,, \label{sdp_balance}
\end{equation}
where $\ell_{\rm right}$ and $X'$ can be determined using the law of sines:
\begin{equation}
\frac{H+L_1\cot\alpha}{\sin\theta}=\frac{\ell_{\rm right}}{\sin\alpha}=\frac{L_1\csc\alpha+X'}{\sin(\pi-\alpha-\theta)}\,,
\end{equation}
which gives
\begin{equation}
\ell_{\rm right}=(H+L_1\cot\alpha)\sin\alpha\csc\theta
\end{equation}
and
\begin{equation}
X'=[L_1\cos(\alpha+\theta)+H\sin(\alpha+\theta)]\csc\theta\,.
\end{equation}

Substituting  $\ell_{\rm right}$ and $X'$ into (\ref{sdp_balance}), we obtain the minimal aspect ratio required for SDP condensation
\begin{equation}
a^{\rm SDP}(\theta,\alpha)=\frac{\sin(\alpha+\theta)+1}{\cos(\alpha+\theta)+\cos\theta}\,. \label{amin_sdp}
\end{equation}
The positivity of $a^{\rm SDP}$ imposes the following constraint
\begin{equation}
\theta+\frac{\alpha}{2}<\frac{\pi}{2}\;\;\;{\rm (SDP)}\,. \label{sdp_angle}
\end{equation}
This condition is equivalent to requiring $\delta p_{\rm SP-SDP}>0$, which implies that the transition pressure $p_{\rm SP-SDP}$  remains below the
saturation pressure $p_{\rm sat}$.

From the other side, the stability of the SDP regime relative to the DP regime at saturation necessitates the condition
\begin{equation}
\alpha+\theta\ge\frac{\pi}{2}\;\;\;{\rm (SDP)}\,, \label{sdp_angle2}
\end{equation}
as demonstrated previously.

\begin{figure}[t]
  \includegraphics[width=0.9\linewidth]{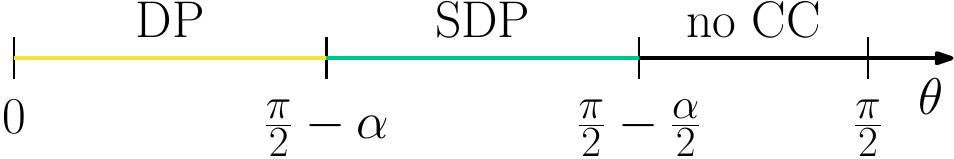}
  \caption{One-dimensional diagram illustrating the condensation state at saturation as a function of Young's contact angle $\theta$, for a fixed opening angle $\alpha$, assuming the system is
    sufficiently large.} \label{dp0_theta}
\end{figure}

\begin{figure}[t]
  \subfloat{\columnwidth}{$\theta<\pi/4$.}{%
  \includegraphics[width=0.9\linewidth]{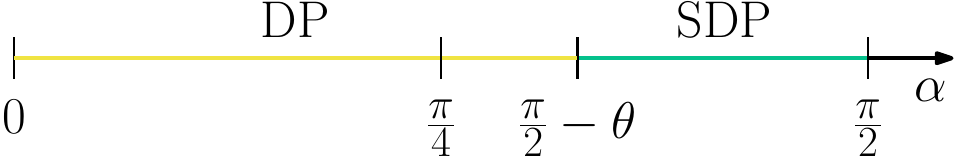}\vspace{2ex}%
  }\\%
  \subfloat{\columnwidth}{$\theta>\pi/4$.}{%
  \vspace{3ex}%
  \includegraphics[width=0.9\linewidth]{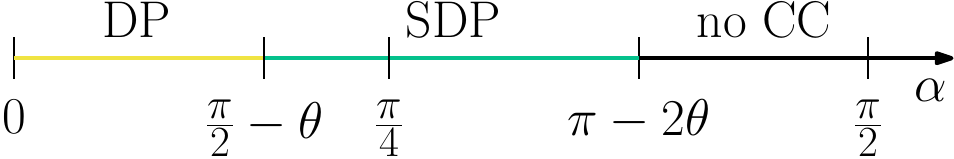}\vspace{2ex}%
  }%
  \caption{One-dimensional diagrams illustrating the condensation state at saturation as a function of the opening angle $\alpha$, for a fixed value of Young's contact angle $\theta$, assuming the
    system is sufficiently large. Separate cases are shown for $\theta < \pi/4$ and $\theta > \pi/4$.} \label{dp0_alpha}
\end{figure}

\subsection{SP regime} \label{sp_limit}

\begin{figure*}[tbh]
  \subfloatflex{$\alpha=30\degree$}{%
    \includegraphics[scale=0.7]{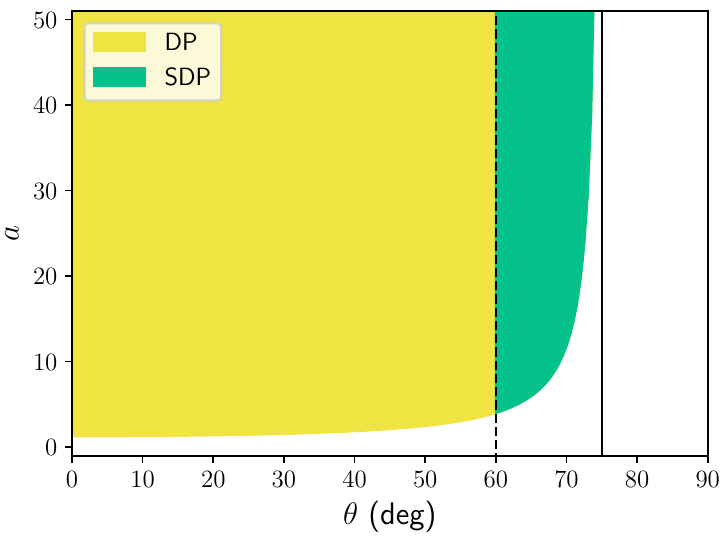}%
  } \hspace*{0.2cm}
  \subfloatflex{$\alpha=60\degree$}{%
    \includegraphics[scale=0.7]{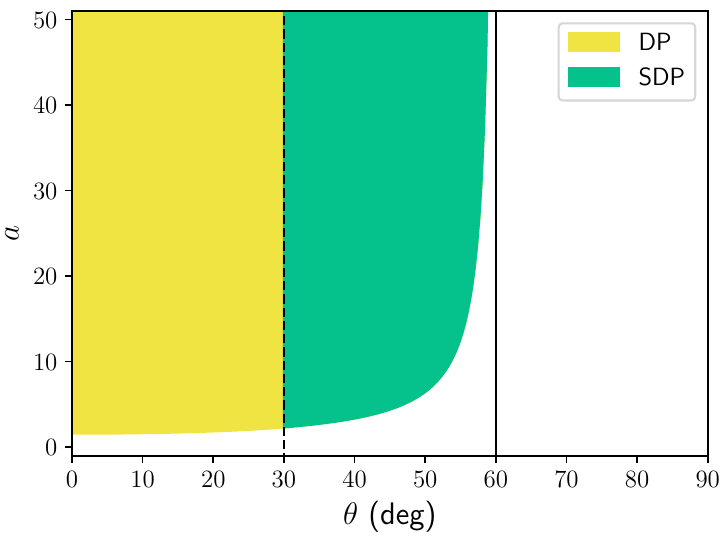}%
  }
  \caption{Two-dimensional extension of the diagram presented in Fig.~\ref{dp0_theta}, incorporating the influence of system size. The dashed vertical line represents the DP/SDP boundary, while the
    solid vertical line marks the SDP condensation limit. } \label{alpha_2d}
\end{figure*}

\begin{figure*}[tbh]
  \subfloatflex{$\theta=30\degree$}{%
    \includegraphics[scale=0.7]{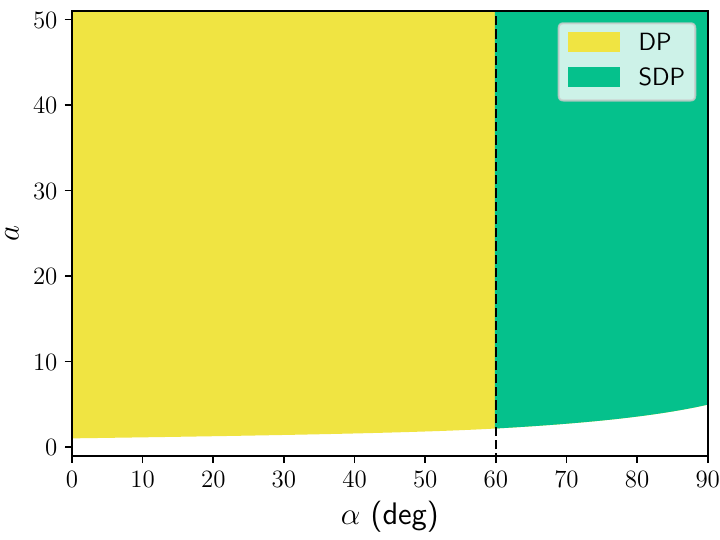}%
  } \hspace*{0.2cm}
  \subfloatflex{$\theta=60\degree$}{%
    \includegraphics[scale=0.7]{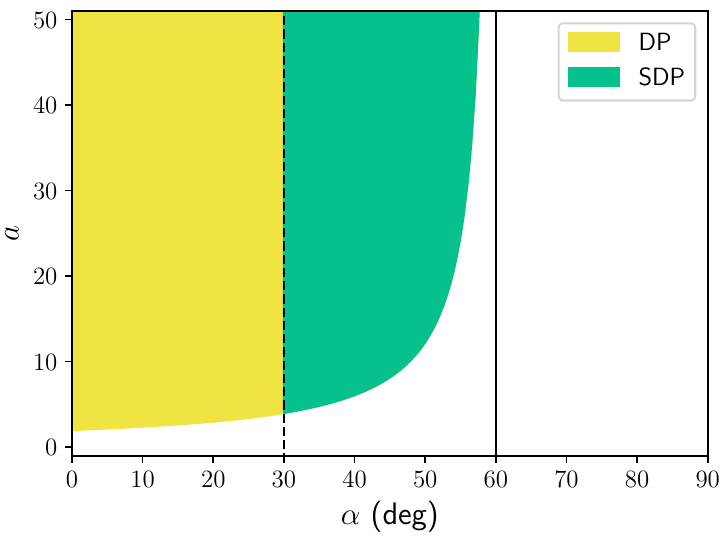}%
  }
  \caption{Two-dimensional extension of the diagram presented in Fig.~\ref{dp0_alpha}, incorporating the effect of system size. The dashed vertical line indicates the DP/SDP boundary, while the solid
    vertical line represents the SDP condensation limit. } \label{theta_2d}
\end{figure*}

The SP state at $\delta p=0$ is possible only under a specific condition: the right meniscus must meet both walls at a contact angle
$\theta=(\pi-\alpha)/2$. Under this condition, the length of the right meniscus is given by
\begin{equation}
\ell_{\rm right}=\csc\theta(L_1\cos\alpha+X\sin\alpha)\,, \label{ell_x}
\end{equation}
where recall $X$ is the size of the bottom wall in contact with the liquid. However, this distance $X$ is arbitrary, as the excess grand potential in
this state
\begin{equation}
\Omega^{\rm ex}/\gamma=\ell_{\rm right}-(X+X')\cos\theta+L_1\,, \label{omex_deloc}
\end{equation}
is unaffected by changing $X$. The invariance follows from Eq.~(\ref{ell_x}), which implies $d\ell_{\rm right}/dX=2\cos\theta$, and the fact that
$dX'/dX=1$. As a result, the right meniscus in this regime becomes delocalized, meaning it can freely move along the walls without any free-energy
cost. However, it can be shown that $\Omega^{\rm ex}>0$ for all $\alpha$, meaning that SP states are always metastable with respect to the
low-density state. Note that the condition $\theta=(\pi-\alpha)/2$ corresponds to the wedge-filling boundary (\ref{filling}) in a linear wedge with
opening angle $\alpha$ determining implicitly the filling temperature $T_f$. In this case, however, the additional free-energy cost due to the
presence of the left meniscus precludes this scenario, at least on the macroscopic scale.

\begin{figure*}[tbh]
  \centering%
  \subfloat{0.66\columnwidth}{$a=1.01$}{%
    \includegraphics[width=\textwidth]{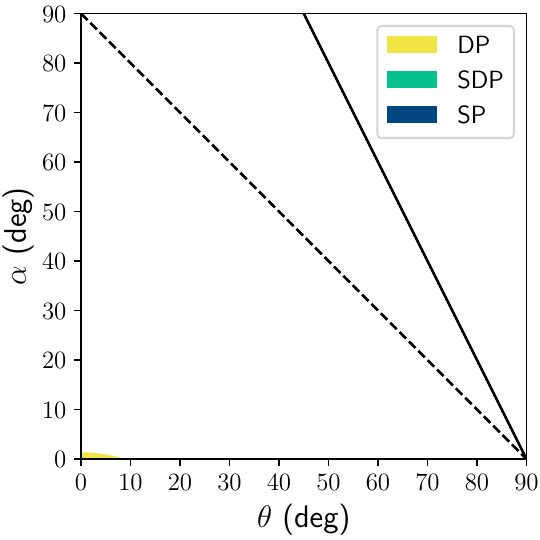}%
  }%
  \hspace{0.02\columnwidth}%
  \subfloat{0.66\columnwidth}{$a=1.5$}{%
    \includegraphics[width=\textwidth]{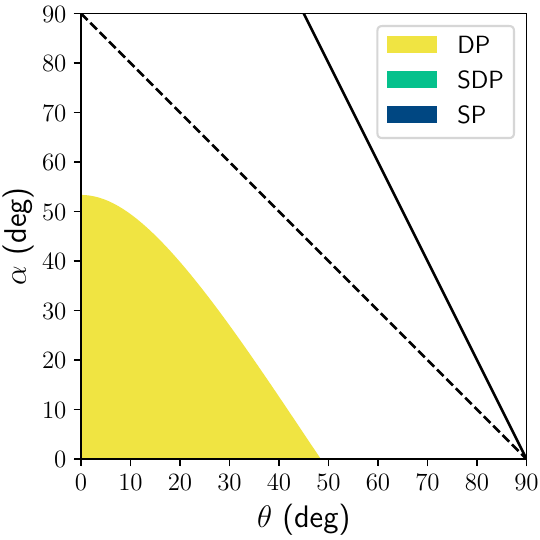}%
  }%
  \hspace{0.02\columnwidth}%
  \subfloat{0.66\columnwidth}{$a=2$}{%
    \includegraphics[width=\textwidth]{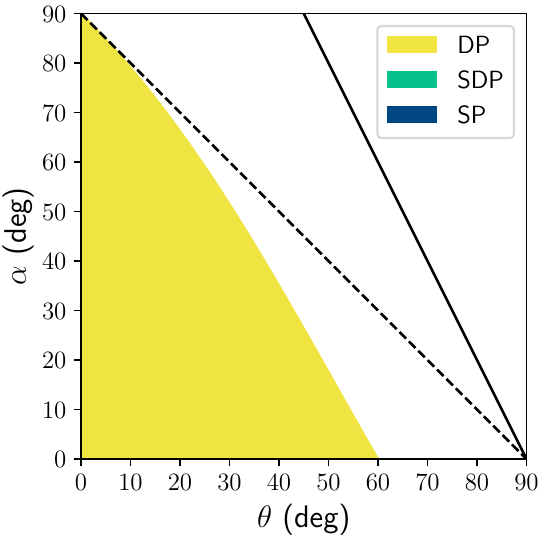}%
  }%
  \caption{Phase diagrams of short systems exhibiting only DP condensation. The boundary of the DP condensation region is denoted by the dashed line, $\alpha + \theta = \pi / 2$. The solid line
    represents the condensation limit, given by $\alpha = \pi - 2\theta$.} \label{pd_short}
\end{figure*}

\vspace*{0.5cm}

In the upshot, for the system to condense, the following condition must be met
\begin{equation}
a>a_{\rm min}(\theta,\alpha)
\end{equation}
where $a_{\rm min}$ depends on the relationship between $\alpha$ and  $\theta$:
\begin{itemize}
  \item If $\alpha+\theta<\pi/2$, then $a_{\rm min}=a^{\rm DP}$ [cf. Eq.~(\ref{amin_sdp})], and the system condenses into the DP state.

  \item  If $\alpha+\theta>\pi/2$ but $\alpha/2+\theta<\pi/2$, then $a_{\rm min}=a^{\rm SDP}$ [cf. Eq.~(\ref{amin_sdp})] and the system condenses into SDP state.

  \item If $\alpha/2+\theta>\pi/2$, the system does not exhibit any type of capillary condensation.

\end{itemize}

Thus, fixing the system geometry and allowing only for variations in $\theta$ (e.g., by changing the temperature), we can summarize the behavior of
the system in the limit $\delta p\to0$ by defining $\theta^-=\pi/2-\alpha$ and $\theta^+=\pi/2-\alpha/2$:
\begin{equation}
\theta\in\begin{cases}
  (0,\theta^-);             & {\rm DP}           \\
  (\theta^-,\theta^+);      & {\rm SDP}          \\
  (\theta^+,\frac{\pi}{2}); & {\rm no\;\;CC}\,,
\end{cases}
\end{equation}
where we emphasize that both DP and SDP condensations require $a>a_{\rm min}(\theta,\alpha)$ for their occurrence. This behavior is illustrated
graphically in Fig.~\ref{dp0_theta}.

Alternatively, we can fix $\theta$ and examine the impact of the upper wall inclination. In this case, two scenarios arise:
\begin{itemize}
  \item  If $\theta>\pi/4$,  there are three possibilities: DP condensation occurs for small values of $\alpha$, SDP condensation occurs for $\pi/2-\theta<\alpha<\pi-2\theta$, and no capillary condensation is possible for
        $\alpha>\pi-2\theta$.

  \item If $\theta<\pi/4$, the system (of sufficient length) will always condense into one of the DP or SDP regimes, as shown in Fig.~\ref{dp0_alpha}.
\end{itemize}

The effect of the system size is further explored in two-dimensional graphs illustrating the equilibrium states in the limit $\delta p\to0$ by
varying  $a$ and $\theta$ for a fixed $\alpha$ (Fig.~\ref{alpha_2d}) and by varying $a$ and $\alpha$ for a fixed $\theta$ (Fig.~\ref{theta_2d}).  In
these diagrams, the DP and SDP regimes are separated by the lines $a^{\rm DP}$ and $a^{\rm SDP}$, which connect smoothly. Below these lines, no
capillary condensation is possible.

\vspace*{0.5cm}

To conclude this section, we summarize the results for the $\delta p\to0$ limit which already reveals some important features of the system phase
behavior:
\begin{enumerate}
  \item  The condition $\alpha/2+\theta<\pi/2$ must be satisfied for any type of capillary condensation to occur.

  \item In the region defined by $\alpha/2+\theta<\pi/2$ and $\alpha+\theta>\pi/2$,
        the equilibrium condensation state belongs to the SDP regime. The following scenarios are then possible:

        \begin{enumerate}

          \item Initial condensation into the SP state, followed by a continuous depinning transition to the SDP state at the pressure given
                by Eq.~(\ref{p_sp_sdp}). This requires that $a>a^{\rm SP}$.

          \item Direct condensation into the SDP state. In this case $a^{\rm SP}>a>a^{\rm SDP}$.

          \item Remaining in the low-density state if $a<a^{\rm SDP}$.

        \end{enumerate}

  \item  DP condensation is only possible when $\alpha+\theta<\pi/2$.
        In this case, all three condensation regimes (DP, SDP, and SP) are accessible, with the DP regime specifically requiring $a>a^{\rm DP}$.

\end{enumerate}



\section{Global phase diagrams}

In this section, we utilize the results from Sec.~\ref{regimes}  to construct the global phase diagrams. These diagrams are plotted in the
$\alpha-\theta$ plane, where each point represents the type of condensation (if any) that occurs for a fixed value of the aspect ratio $a$. The
discussion is divided into four parts, corresponding to distinct intervals of $a$, each exhibiting qualitatively different behavior.

\subsection{Short systems}

\begin{figure*}[tbh]
  \centering%
  \subfloat{0.66\columnwidth}{$a=2.5$}{%
    \includegraphics[width=\textwidth]{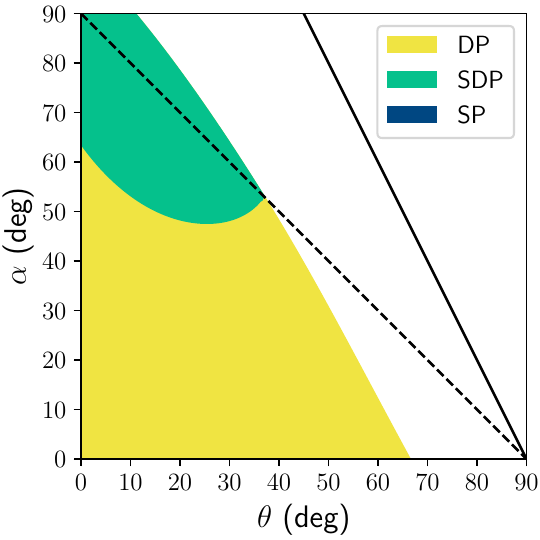}%
  }%
  \hspace{0.02\columnwidth}%
  \subfloat{0.66\columnwidth}{$a=3$}{%
    \includegraphics[width=\textwidth]{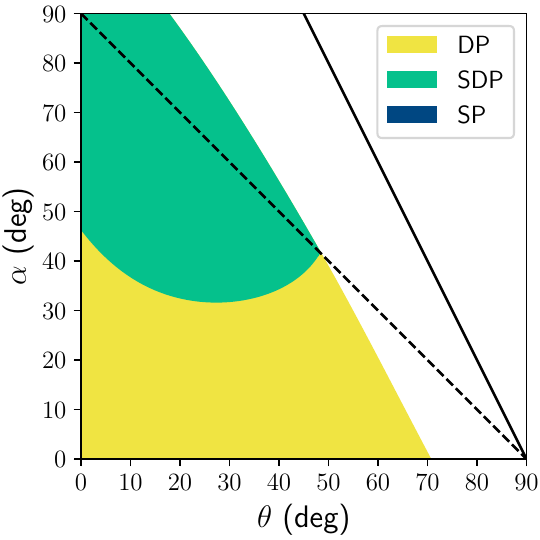}%
  }%
  \hspace{0.02\columnwidth}%
  \subfloat{0.66\columnwidth}{$a=4$}{%
    \includegraphics[width=\textwidth]{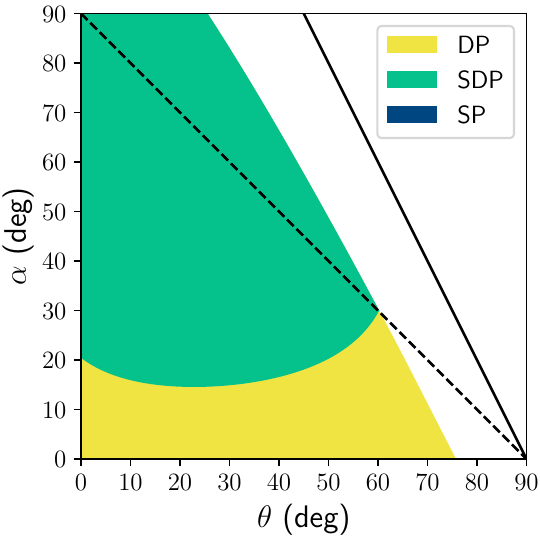}%
  }%
  \caption{Phase diagrams of moderately long systems that may undergo both DP  and SDP condensations. The boundary for DP condensation is denoted by
    the dashed line, $\alpha= \pi/2-\theta$, while the solid line represents the condensation limit, $\alpha = \pi - 2\theta$. The DP/SDP boundary, given
    by Eq.~(\ref{dp_sdp}), terminates at the dashed line, corresponding to the point $\theta^* = \cos^{-1}(2/a)$ [see Eq.~(\ref{theta_ast})].}
  \label{pd_med}
\end{figure*}

We start by analyzing systems with small aspect ratios, specifically within the interval $1<a<2$. In this case, the system is too short to undergo
any condensation except for DP type. The lower bound of this interval (and the lowest bound for condensation) is given by the global minimum of
$a_{\rm min}$, which occurs at $\alpha=\theta=0$. In this configuration, the system reduces to a finite slit formed of completely wet walls. The
upper limit of this interval corresponds to the smallest $a$ where SDP condensation replaces DP condensation. This occurs when $a^{\rm DP}=a^{\rm
SDP}$, a condition satisfied at the limiting case of  $\alpha=\pi/2$ with $\theta=0$. This marks the minimal $a$ for which $a^{\rm
DP}(\theta,\alpha)$ reaches the DP limit, defined by $\alpha+\theta=\pi/2$.

For $1<a<2$, the area of the DP region in the $\alpha$-$\theta$ plane progressively  grows with increasing $a$, as demonstrated in
Fig.~\ref{pd_short}. These diagrams illustrate how the DP regime expands, with its boundaries governed by the $a^{\rm DP}$ curve and the limiting
angle condition $\alpha+\theta=\pi/2$.

\subsection{Moderately long systems}

We now examine systems with aspect ratios in the interval  $2<a<a^{\rm SP}_{\rm min}$, where two distinct condensation regimes are possible: DP and
SDP. The upper bound of this interval, $a^{\rm SP}_{\rm min}$, corresponds to the onset of SP condensation, which will be discussed in the subsequent
paragraph.

In this interval, $a^{\rm DP}$ intersects the line  $\alpha+\theta=\pi/2$ at the transition point $(\theta^*,\alpha^*)$, defined by the condition:
\begin{equation}
\cos\theta^*=\frac{2}{a}\,, \label{theta_ast}
\end{equation}
as follows from Eq.~(\ref{amin_dp}). At this point the $a^{\rm DP}$ curve  which determines the condensation boundary for  $\alpha<\alpha^*$,
connects with $a^{\rm SDP}$, the corresponding boundary for $\alpha>\alpha^*$. The $a^{\rm SDP}$ curve further extends beyond the global DP limit,
$\alpha=\pi/2-\theta$, but is of course bounded by the condensation limit $\alpha=\pi-2\theta$.

The boundary between the DP and SDP regimes, say $\tilde{\alpha}(\theta,a)$, can be determined by substituting from Eqs.(\ref{dp_theta2}) and
(\ref{p_sp_sdp}) into Eq.~(\ref{cc_dp}). This yields:
\begin{multline}
  \hspace{-\multlinegap}
  \frac{1}{2} \, {\left(\pi - \tilde{\alpha} - \theta - \theta_{1}\right)} {\left(a \tan\tilde{\alpha} + 1\right)}
  \sec\left(\tilde{\alpha} + \theta\right) \\
  + \frac{{\left(a \tan\tilde{\alpha} + 2\right)} a \cos\left(\tilde{\alpha} + \theta\right)}{a \tan\tilde{\alpha} + 1} + \frac{1}{2} \, \sqrt{-\frac{\cos\left(\tilde{\alpha} + \theta\right)^{2}}{{\left(a
          \tan\tilde{\alpha} + 1\right)}^{2}} + 1} \\
  -a {\left(\sec\tilde{\alpha} + 1\right)} \cos\theta
  + \frac{1}{2} \, {\left(a \tan\tilde{\alpha} + 1\right)} \sin\left(\tilde{\alpha}+\theta\right) =0\,, \label{dp_sdp}
\end{multline}
where $\theta_1=\cos^{-1}\left[\cos(\tilde{\alpha}+\theta)/(1+a\tan\tilde{\alpha})\right]$.   This equation applies to the range of $\theta$ between $0$ and $\theta^*$.

The phase behavior for moderately long systems is illustrated in Fig.~\ref{pd_med}. With increasing $a$, the transition point $(\theta^*,\alpha^*)$
shifts from the upper left corner along the $\alpha+\theta=\pi/2$ line.  This reflects the gradual expansion of the SDP region at the expense of the
DP region. This culminates at the maximum aspect ratio within this category, $a=a^{\rm SP}_{\rm min}$, where a new condensation regime -- the SP
regime -- emerges.

\subsection{Long Systems}

\begin{figure*}[pth]
  \subfloatflex{Dependence $a^{\rm SP}(\theta)$.}{%
    \includegraphics[width=0.98\columnwidth]{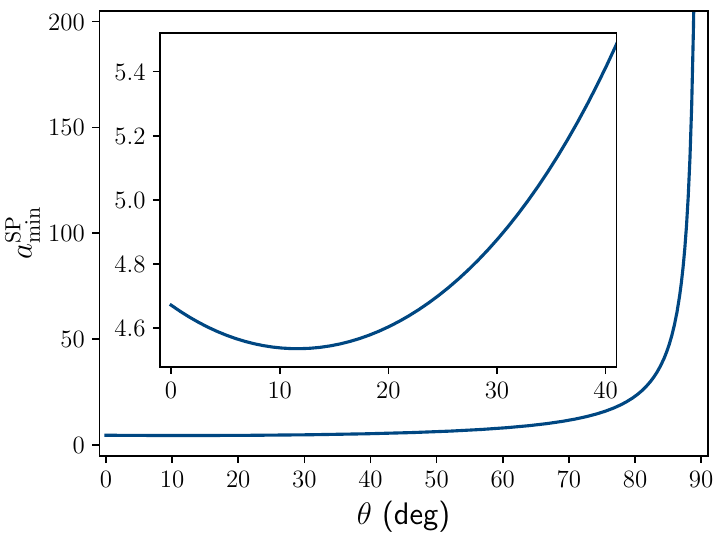}%
  } \hspace*{0.2cm}
  \subfloatflex{Dependence of the
  opening angle $\alpha$ corresponding to $a^{\rm SP}(\theta)$ on $\theta$. This curve is part of the phase diagrams shown in Figs.~\ref{pd_long} and \ref{pd_huge}.}{%
    \includegraphics[width=0.98\columnwidth]{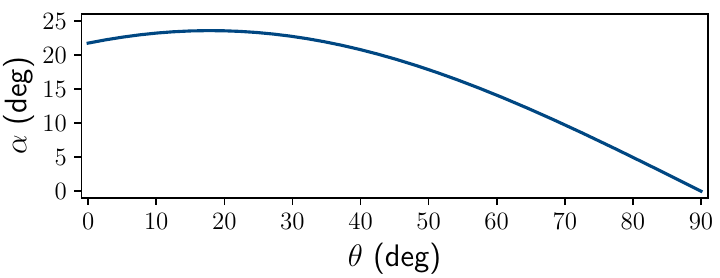}\vspace{2em}%
  }%
  \caption{The smallest aspect ratio, $a^{\rm SP}$, required for SP condensation on $\theta$, obtained by numerically solving Eq.~(\ref{eq:dRdalpha_theta}). The curve exhibits a
    minimum at $\theta\approx11.6\degree$, which corresponds to $a^{\rm SP}_{\rm min}\approx4.54$. This value separates
    ``moderately long systems'' from ``long systems''.} \label{fig_sp_theta}
\end{figure*}

\begin{figure*}[pth]
  \subfloat{0.66\columnwidth}{$a=4.54$}{%
    \includegraphics[width=0.98\textwidth]{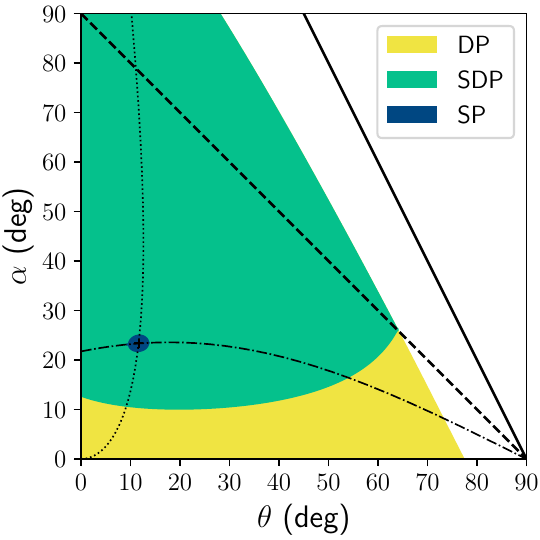}%
  }%
  \hspace{0.01\columnwidth}%
  \subfloat{0.66\columnwidth}{$a=4.6$}{%
    \includegraphics[width=0.98\textwidth]{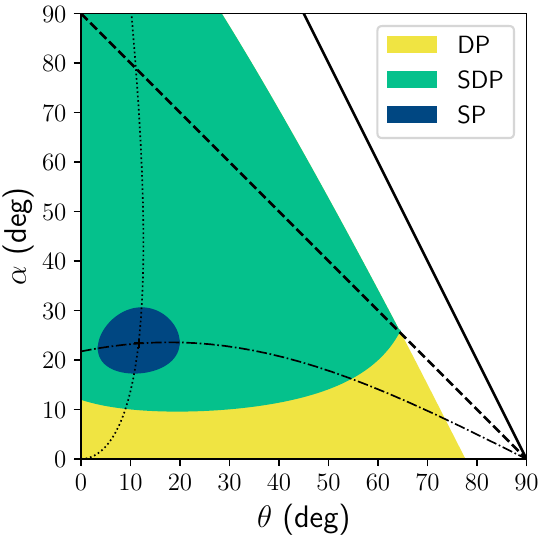}%
  }%
  \hspace{0.01\columnwidth}%
  \subfloat{0.66\columnwidth}{$a=4.67$}{%
    \includegraphics[width=0.98\textwidth]{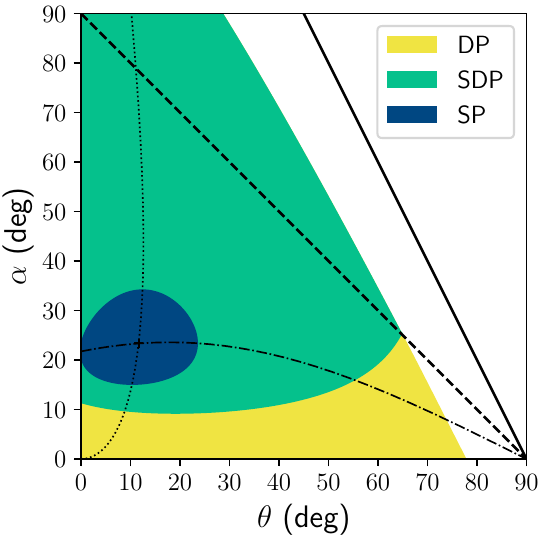}%
  }%

  \subfloat{0.66\columnwidth}{$a=4.7$}{%
    \includegraphics[width=0.98\textwidth]{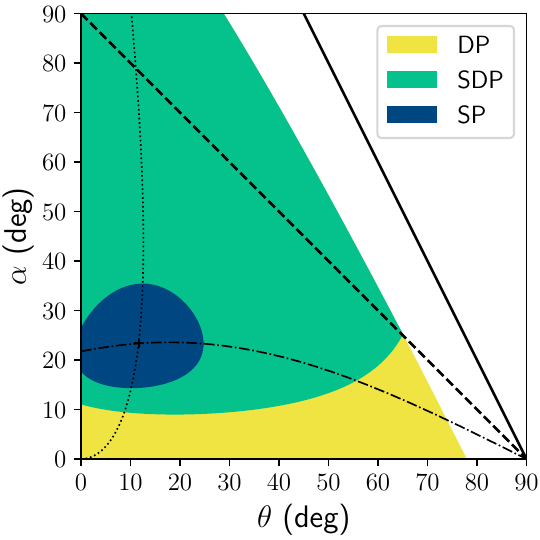}%
  }%
  \hspace{0.01\columnwidth}%
  \subfloat{0.66\columnwidth}{$a=5$}{%
    \includegraphics[width=0.98\textwidth]{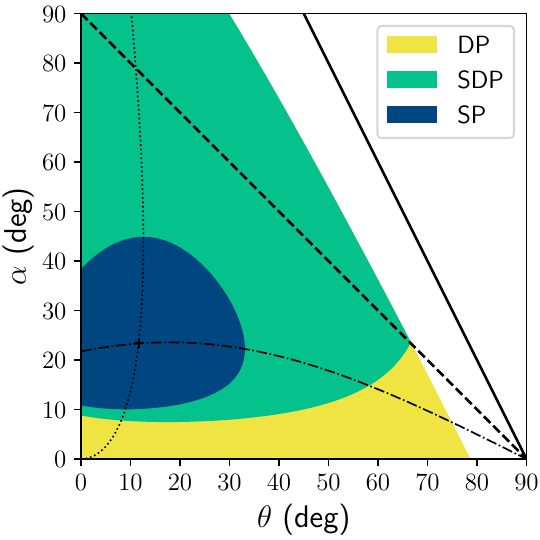}%
  }%
  \hspace{0.01\columnwidth}%
  \subfloat{0.66\columnwidth}{$a=7$}{%
    \includegraphics[width=0.98\textwidth]{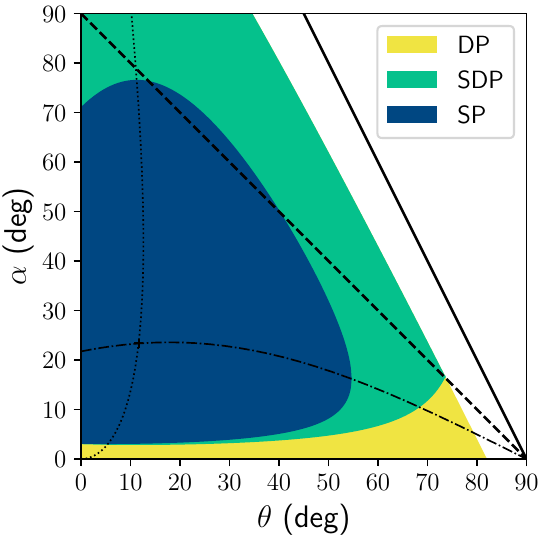}%
  }%
  \caption{Phase diagrams of long systems exhibiting all types of condensation: DP, SDP, and SP. The boundary for DP condensation is denoted by the dashed line, $\alpha = \pi / 2 - \theta$, while the solid line
    represents the condensation limit, $\alpha = \pi - 2\theta$. The dash-dotted line shows the loci of $\alpha$ corresponding to $a^{\rm SP}(\theta)$,
    and the dotted line represents the loci of $\theta$ corresponding to $a^{\rm SP}(\alpha)$.} \label{pd_long}
\end{figure*}

For sufficiently long systems, all three condensed phases -- DP, SDP, and SP -- can occur.  Specifically, to allow for an SP condensation the
condition $H>X$ must be satisfied. To determine this requirement, we proceed as follows. Let us find the minimum of $X$ as a function of $\alpha$,
while keeping $\theta$ constant, i.e., we solve the condition:
\begin{equation}
\left(\pdiff{X}{\alpha}\right)_{\theta=\mathrm{const.}}=0\,, \label{Xalpha}
\end{equation}
subject to the constraint  $\Omega^\mathrm{ex}_\mathrm{SP}=0$.

In solving Eq.~(\ref{Xalpha}), one should account for the fact that $R$  is itself  a function of $\alpha$ and its derivative with respect to
$\alpha$ can be expressed as
\begin{equation}
\left(\pdiff{R}{\alpha}\right)_{\theta=\mathrm{const.},\Omega^\mathrm{SP}=0}=-\frac{\left(\pdiff{\Omega^\mathrm{SP}}{\alpha}\right)_{\theta,R}}
{\left(\pdiff{\Omega^\mathrm{SP}}{R}\right)_{\theta,\alpha}}\,. \label{eq:dRdalpha_theta}
\end{equation}
The solution of Eq.~(\ref{Xalpha}) yields the minimum aspect ratio $a^{\rm SP}(\theta)$ required for SP condensation, as a function of $\theta$. The
dependence, depicted in Fig.~\ref{fig_sp_theta}a, is non-monotonic and exhibits a minimum at $\theta\approx11.6\degree$, for which $a^{\rm
      SP}\approx4.54$, corresponding to the sought aspect ratio $a^{\rm SP}_{\rm min}$. The corresponding opening angle, $\alpha$, as a function of
$\theta$, is shown in Fig.~\ref{fig_sp_theta}b.

The phase behavior of systems with $a>a^{\rm SP}_{\rm min}$ is illustrated in Fig.~\ref{pd_long}. Initially,  for $a$ slightly greater than $a^{\rm
SP}_{\rm min}$, the SP region is enclosed by a loop in the $(\alpha,\theta)$ plane. This loop spans between the contact angles $\theta^->0$ and
$\theta^+$, corresponding to the intersections of $a$ with the $a^{\rm SP}$ curve in Fig.~\ref{fig_sp_theta}a. However,  for $a\approx4.67$ the SP
loop reaches the vertical axis, allowing SP condensation even for completely wet walls ($\theta=0$).  As $a$ increases further, the SP region grows
rapidly at the expense of the SDP regime. Notably, the SP region is entirely embedded within the SDP region, as no direct phase boundary exists
between SP and DP phases. However, the band separating SP and DP regions may become difficult to discern visually.

\subsection{Very long systems}

\begin{figure}[t!]
  \subfloat{\linewidth}{Dependence $a^\mathrm{SP}(\alpha)$.}{%
   \includegraphics[width=\linewidth]{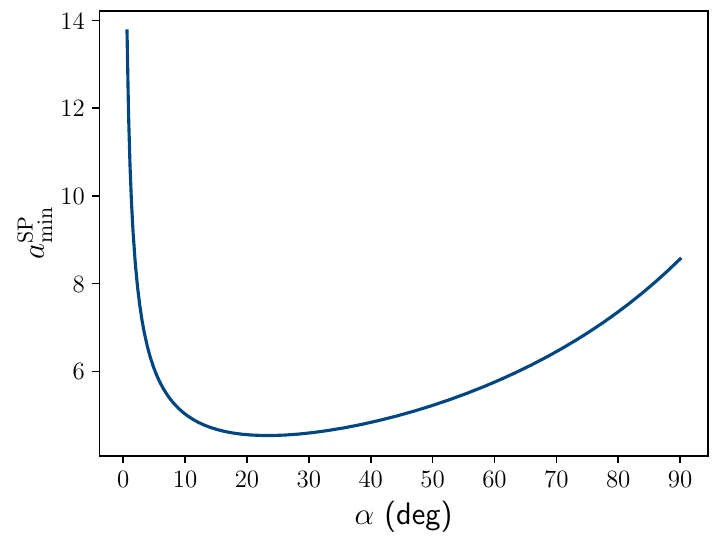}%
  }\\[0.5em]
  \subfloat{\linewidth}{Dependence of $\theta$ corresponding to $a^{\rm SP}(\alpha)$ on $\alpha$. This curve is part of the phase diagrams shown in Figs.~\ref{pd_long} and \ref{pd_huge}.}{%
  \hspace{0.03\linewidth}\includegraphics[width=0.97\linewidth]{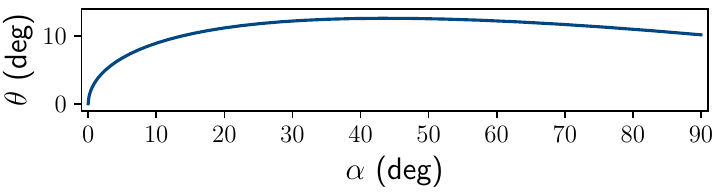}%
  }%
  \caption{The smallest aspect ratio, $a^{\rm SP}$,  allowing for SP condensation on $\alpha$, obtained by numerically solving
    Eq.~(\ref{eq:dXdtheta_alpha}).} \label{fig_sp_alpha}
\end{figure}

\begin{figure*}[tbh!]
  \centering%
  \subfloat{0.66\columnwidth}{$a=8.56$}{%
    \includegraphics[width=\textwidth]{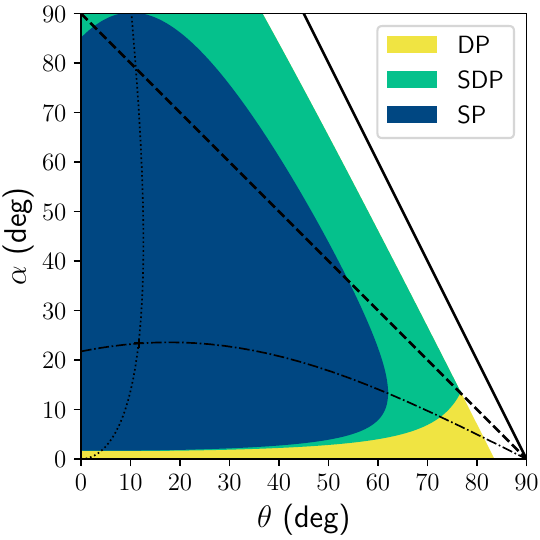}%
  }%
  \hspace{0.02\columnwidth}%
  \subfloat{0.66\columnwidth}{$a=8.6$}{%
    \includegraphics[width=\textwidth]{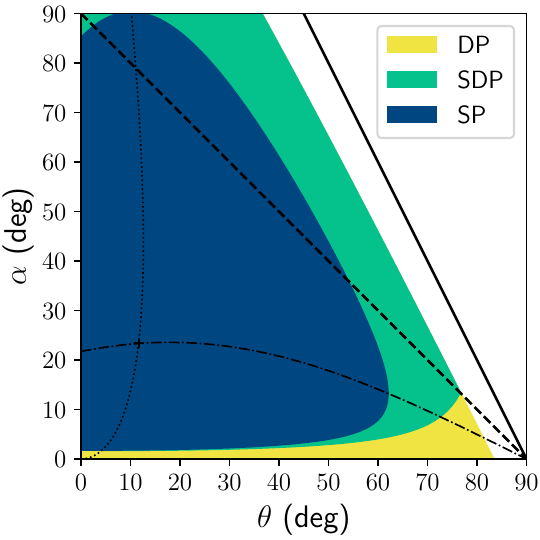}%
  }%
  \hspace{0.02\columnwidth}%
  \subfloat{0.66\columnwidth}{$a=10$}{%
    \includegraphics[width=\textwidth]{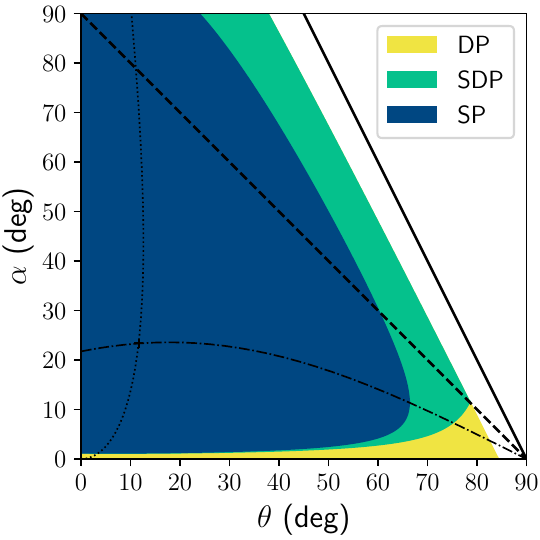}%
  }%
  \caption{Phase diagrams of very long systems, characterized by an extension of the SP region up to the limit of $\alpha = \pi/2$. The boundary for DP
    condensation is denoted by the dashed line, $\alpha = \pi/2 - \theta$, while the solid line represents the condensation limit, $\alpha = \pi -
      2\theta$. The first two plots correspond to the case $a < a_{\rm dis}$, while the last plot corresponds to $a> a_{\rm dis}$, where $a_{\rm dis}
      \approx 9.32$ is given by Eq.~(\ref{adis}). Also shown are the loci of $\alpha$ corresponding to $a^{\rm SP}(\theta)$ (dash-dotted line) and the loci
    of $\theta$ corresponding to $a^{\rm SP}(\alpha)$ (dotted line).} \label{pd_huge}
\end{figure*}

Finally, we turn to the class of very long systems, characterized by the occurrence of SP condensation even in the limit  $\alpha\to\pi/2$ (for
certain values of $\theta$). To determine the minimal aspect ratio where this behavior emerges, we follow a similar approach to the previous case by
solving
\begin{equation}
  \left(\pdiff{X}{\theta}\right)_{\alpha=\mathrm{const.}}=0
  \label{eq:dXdtheta_alpha}
\end{equation}
subject to the condition $\Omega^\mathrm{SP}=0$. Here, the radius $R$ is treated as a function of $\theta$, and its derivative is obtained as
\begin{equation}
  \left(\pdiff{R}{\theta}\right)_{\alpha=\mathrm{const.}} =
  -\frac{\left(\pdiff{\Omega^\mathrm{SP}(R,\alpha,\theta)}{\theta}\right)_{\alpha,R}}{\left(\pdiff{\Omega^\mathrm{SP}(R,\alpha,\theta)}{R}\right)_{\theta,\alpha}}\,.
\end{equation}

The solution of Eq.~(\ref{eq:dXdtheta_alpha}) yields the minimum value of $a^{\rm SP}$, as well as the corresponding value of $\theta$, required for
SP condensation as a function of $\alpha$. This dependence is shown in Fig.~\ref{fig_sp_alpha}. The minimum of $a^{\rm SP}(\alpha)$ occurs at
$\alpha\approx23\degree$,  corresponding to $a^{\rm SP}\approx4.54$ and $\theta\approx11.6\degree$, consistent with earlier findings.

By substituting $\alpha=\pi/2$, we find $a^{\rm SP}_{\pi/2}\equiv a^{\rm SP}(\pi/2)\approx8.56$, which defines  the aspect ratio separating the last
two classes of systems.

Within the current class, the phase behavior can be further divided into two distinct cases:

\begin{enumerate}

  \item \emph{Disjoint SDP Regions ($a<a_{\rm dis}$)}:
        For aspect ratios smaller than $a_{\rm dis}$, the SP region separates the SDP region into two disconnected subsets in the $(\alpha,\theta)$ plane.

  \item \emph{Connected SDP Region ($a>a_{\rm dis}$)}:
        When the aspect ratio exceeds $a_{\rm dis}$, there is only one  SDP region, forming a connected set.

\end{enumerate}

The  aspect ratio $a_{\rm dis}$ can be determined by substituting $\theta=0$, $\alpha=\pi/2$ and $R/L_1=a_{\rm dis}$   (from the geometry) into
Eq.~(\ref{omex_sp})
\begin{equation}
a_{\rm dis}\left[\frac{3}{4}\pi-1-\cos^{-1}\left(\frac{1}{2a_{\rm dis}}\right)\right]+\frac{\sqrt{4a_{\rm dis}^{2}-1}}{4a_{\rm
      dis}}=-1\,. \label{adis}
\end{equation}
Solving this equation numerically gives $a_{\rm dis}\approx9.32$. The phase behavior of very long systems is illustrated in Fig.~\ref{pd_huge} for
three representative cases. These diagrams reveal the rapid growth of the SP region both below and above $a_{\rm dis}$.

\subsection{Equilibrium condensation profiles and their evolution}

\begin{figure*}[ptbh]
  \noindent\subfloat{0.245\textwidth}{$a=1.20$}{%
    \includegraphics[scale=0.6]{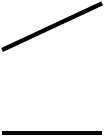}%
    \vspace{0.7em}%
  }\hspace{0.006\textwidth}%
  \subfloat{0.245\textwidth}{$a=1.25$}{%
    \includegraphics[scale=0.6]{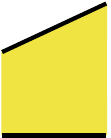}%
    \vspace{0.7em}%
  }\hspace{0.006\textwidth}%
  \subfloat{0.245\textwidth}{$a=2.00$}{%
    \includegraphics[scale=0.6]{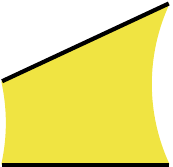}%
    \vspace{0.7em}%
  }\hspace{0.006\textwidth}%
  \subfloat{0.245\textwidth}{$a=3.00$}{%
    \includegraphics[scale=0.6]{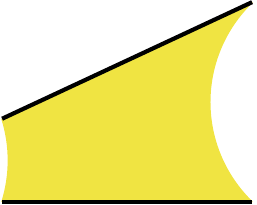}%
    \vspace{0.7em}%
  }\vskip1em
  \subfloat{0.245\textwidth}{$a=3.50$}{%
    \includegraphics[scale=0.6]{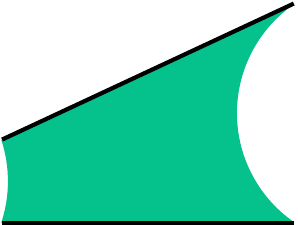}%
    \vspace{0.7em}%
  }\hspace{0.006\textwidth}%
  \subfloat{0.245\textwidth}{$a=4.00$}{%
    \includegraphics[scale=0.6]{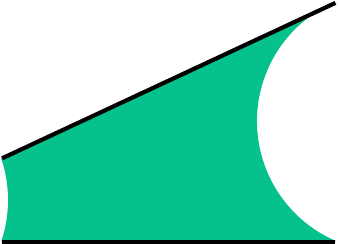}%
    \vspace{0.7em}%
  }\hspace{0.006\textwidth}%
  \subfloat{0.245\textwidth}{$a=4.55$}{%
    \includegraphics[scale=0.6]{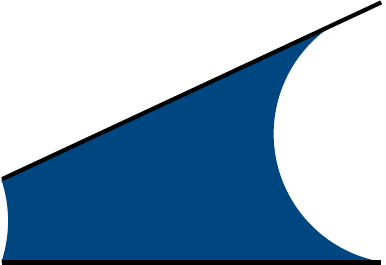}%
    \vspace{0.7em}%
  }\hspace{0.006\textwidth}%
  \subfloat{0.245\textwidth}{$a=5.05$}{%
    \includegraphics[scale=0.6]{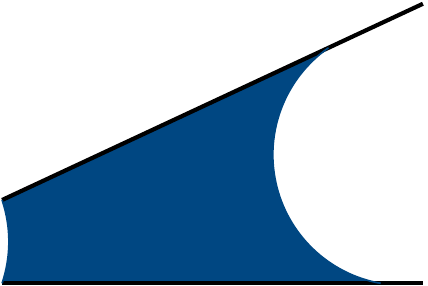}%
    \vspace{0.7em}%
  }%
  \caption{A sequence of equilibrium condensed states of an open wedge with $\alpha = 25^\circ$ and $\theta = 10^\circ$, illustrating the change in the type of condensation as a function of the aspect ratio $a$. 
  All the condensation states are in equilibrium with a low-density (gas-like) state.} \label{states_a}
\end{figure*}

\begin{figure*}[ptbh]
  \subfloatflex{$\theta=0\degree$}{%
    \includegraphics[scale=0.4]{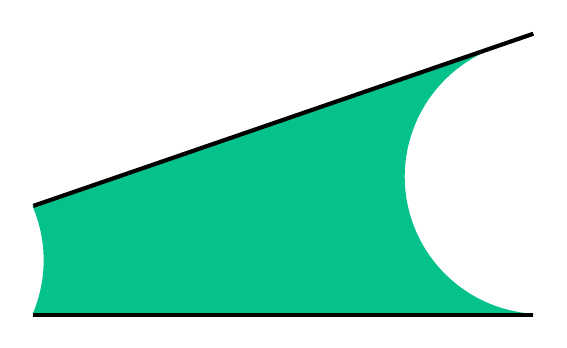}%
  } \hspace*{0.2cm}
  \subfloatflex{$\theta=5\degree$}{%
    \includegraphics[scale=0.4]{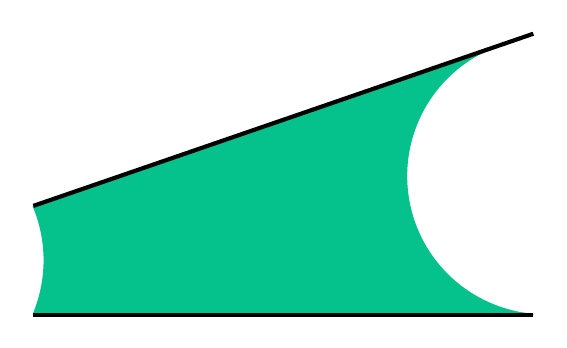}%
  }\hspace*{0.2cm}
  \subfloatflex{$\theta=10\degree$}{%
    \includegraphics[scale=0.4]{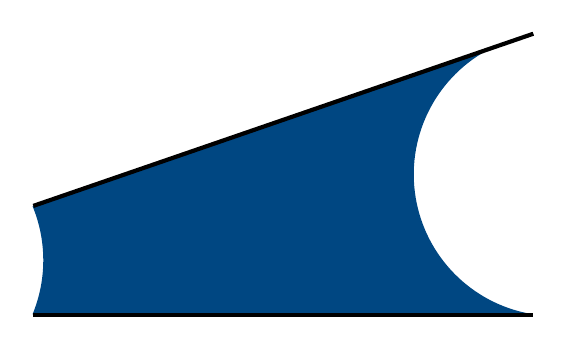}%
  }\hspace*{0.2cm}
  \subfloatflex{$\theta=15\degree$}{%
    \includegraphics[scale=0.4]{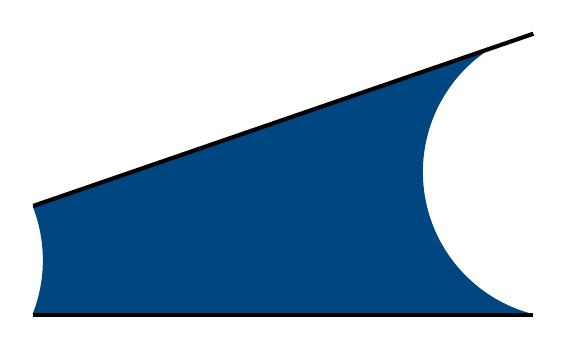}%
  }\vskip1em
  \subfloatflex{$\theta=20\degree$}{%
    \includegraphics[scale=0.4]{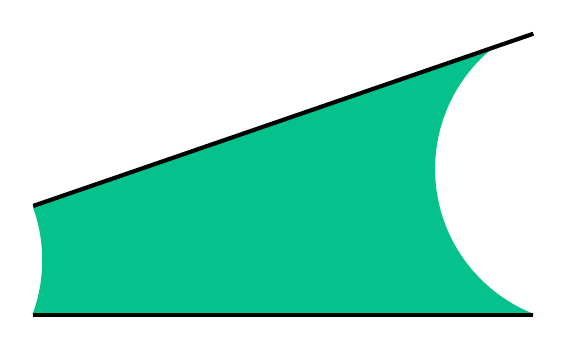}%
  }\hspace*{0.2cm}
  \subfloatflex{$\theta=40\degree$}{%
    \includegraphics[scale=0.4]{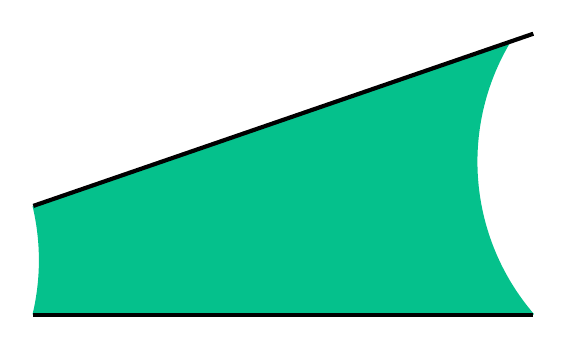}%
  }\hspace*{0.2cm}
  \subfloatflex{$\theta=60\degree$}{%
    \includegraphics[scale=0.4]{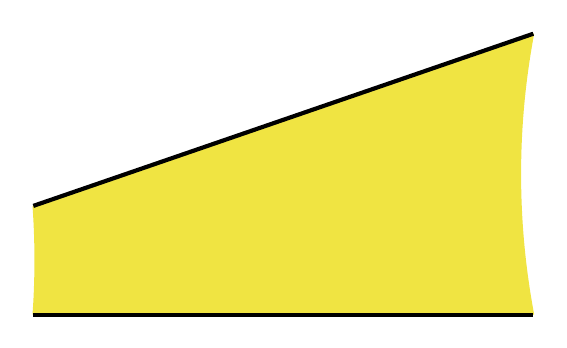}%
  }\hspace*{0.2cm}
  \subfloatflex{$\theta=67\degree$}{%
    \includegraphics[scale=0.4]{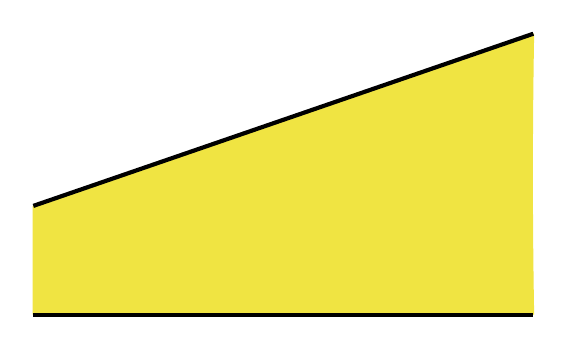}%
  }%
  \caption{A sequence of equilibrium condensed states of an open wedge with $a = 4.6$ and $\alpha = 19^\circ$, illustrating the change in the type of condensation as a function of $\theta$. 
  All the condensation states are in equilibrium with a low-density (gas-like) state.} \label{states_theta}
\end{figure*}

\begin{figure*}[ptbh]
  \noindent\subfloat{0.16\textwidth}{$\alpha=5\degree$}{%
    \includegraphics[width=\textwidth,trim={0 0 0 20cm},clip]{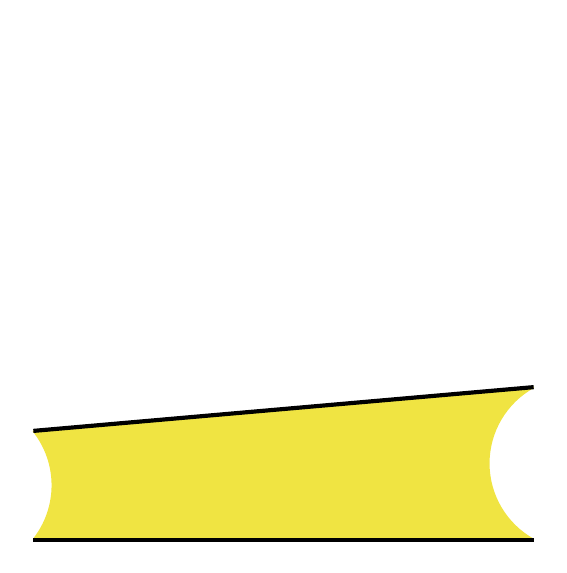}%
    \label{fig:state-alpha-05}%
  }\hspace{0.008\textwidth}%
  \subfloat{0.16\textwidth}{$\alpha=10\degree$}{%
    \includegraphics[width=\textwidth,trim={0 0 0 20cm},clip]{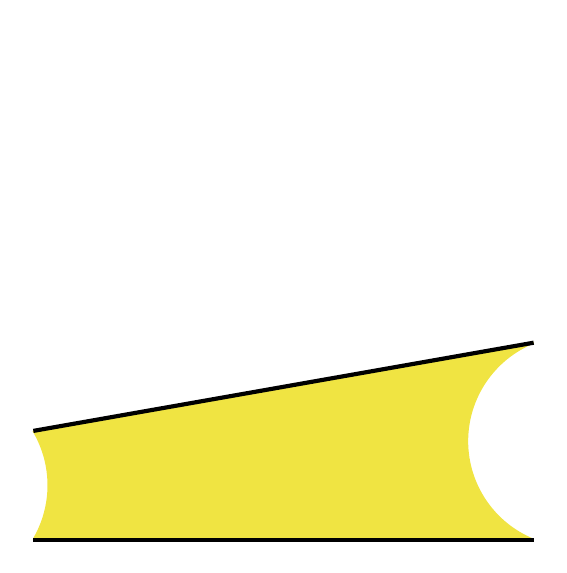}%
    \label{fig:state-alpha-10}%
  }\hspace{0.008\textwidth}%
  \subfloat{0.16\textwidth}{$\alpha=15\degree$}{%
    \includegraphics[width=\textwidth,trim={0 0 0 20cm},clip]{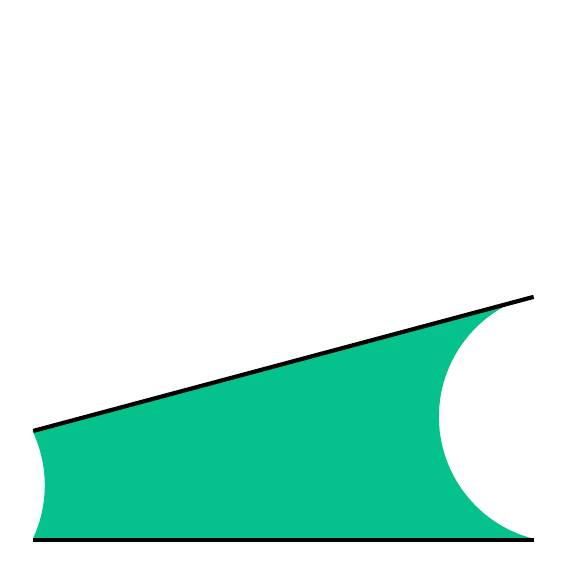}%
    \label{fig:state-alpha-15}%
  }\hspace{0.008\textwidth}%
  \subfloat{0.16\textwidth}{$\alpha=20\degree$}{%
    \includegraphics[width=\textwidth,trim={0 0 0 20cm},clip]{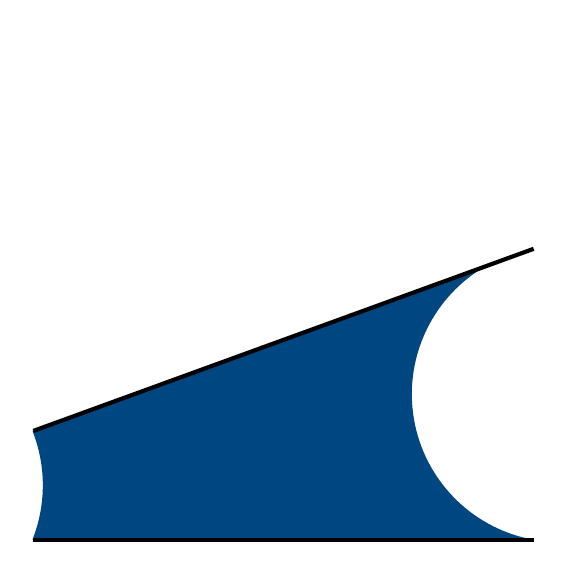}%
    \label{fig:state-alpha-20}%
  }\hspace{0.008\textwidth}%
  \subfloat{0.16\textwidth}{$\alpha=30\degree$}{%
    \includegraphics[width=\textwidth,trim={0 0 0 20cm},clip]{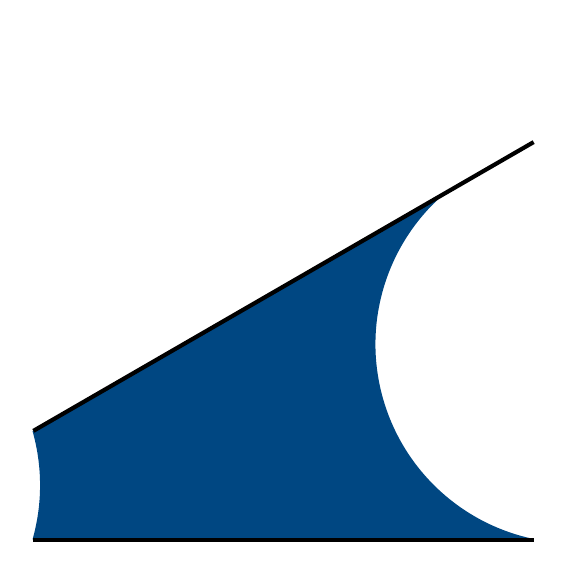}%
    \label{fig:state-alpha-30}%
  }\hspace{0.008\textwidth}%
  \subfloat{0.16\textwidth}{$\alpha=40\degree$}{%
    \includegraphics[width=\textwidth,trim={0 0 0 20cm},clip]{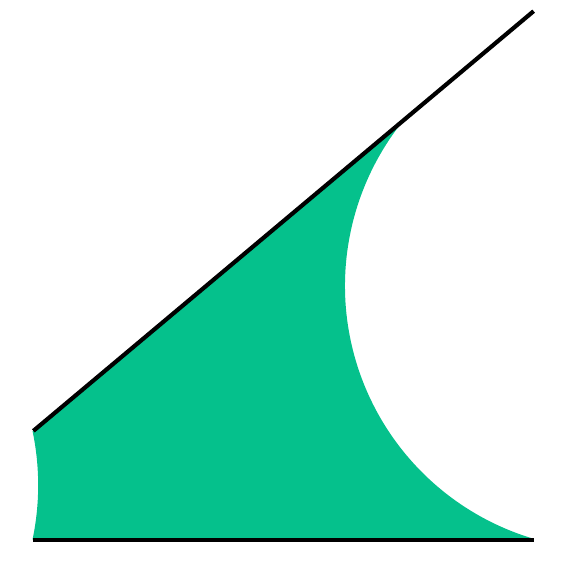}%
    \label{fig:state-alpha-40}%
  }%
  \caption{A sequence of equilibrium condensed states of an open wedge with $a = 4.6$ and $\theta = 12^\circ$, illustrating how the type of condensation varies with $\alpha$. 
  All the condensation states are in equilibrium with a low-density (gas-like) state.} \label{states_alpha}
\end{figure*}

To conclude this section, we present three illustrative sequences of equilibrium states that show how the system evolves as one of the three
parameters -- aspect ratio $a$, contact angle $\theta$, or opening angle $\alpha$ -- is varied while the others are kept fixed.

\subsubsection{Varying the aspect ratio $a$}

We begin by examining the effect of increasing the aspect ratio $a$, fixing the contact and opening angles to $\theta=10\degree$ and
$\alpha=25\degree$. The  equilibrium states corresponding to various cuts of a linear wedge are displayed in Fig.~\ref{states_a}. In order to embrace
the whole variety of possible configurations,  we  begin with a small aspect ratio of $a=1.2$,  slightly below $a_{\rm min}\approx1.246$, where the
system remains in a low-density state even at saturation, as  $\alpha+\theta<\pi/2$ implies $a_{\rm min}=a^{\rm DP}$. At $a=1.25$, just above $a_{\rm
min}$, DP condensation occurs very close to saturation pressure, evident from nearly flat menisci. As the aspect ratio increases, the system still
exhibits only DP condensation but progressively further away from saturation. For higher values of the aspect ratio (such as $a=3.5$ and $a=4$), DP
condensation is preceded by SDP condensation, consistent with the behavior of moderately long systems. Finally, when  $a>a^{\rm SP}_{\rm min}$, SP
condensation becomes possible. Beyond this threshold, further increase in $a$ no longer affects the system  phase behavior.

\subsubsection{Varying the contact angle $\theta$}

Next, we fix $a=4.6$ and $\alpha=19\degree$ to illustrate the effect of varying $\theta$ (e.g., by changing the temperature), see
Fig.~\ref{states_theta}. This sequence corresponds to a horizontal line at $\alpha=19\degree$ in Fig.~\ref{pd_long}(b) for  long systems. Thus, for
small values of $\theta$, the system is initially in SDP regime but at moderate $\theta$, SP condensation takes over. Further increasing $\theta$
reveals a reentrance phenomenon associated with the recurrence of the SDP condensation. At even higher $\theta$, the condensation pressure approaches
saturation, favoring DP condensation since $\alpha<\alpha^*$.  Above $\theta\approx70\degree$, the free-energy gain from the wall-liquid contact is
already too low for the system to condense.

\subsubsection{Varying the opening angle $\alpha$}

Finally, we fix $a=4.6$ and $\theta=12\degree$ to show the impact of changing $\alpha$, see Fig.~\ref{states_alpha}. In this case, the sequence
corresponds to a vertical line in Fig.~\ref{pd_long}(b). For small $\alpha$, DP condensation dominates, as expected for narrow wedges. As $\alpha$
increases, SDP condensation takes over, consistent with the condition $\theta<\theta^*$. By further increasing $\alpha$ the edge contact angle
$\theta_2$ decreases and for $\alpha\approx20\degree$, $\theta_2=\theta$, meaning that the meniscus detaches from  the bottom wall by adopting the SP
state. The right meniscus then recedes inside the wedge before returning back to the opening, when  another reentrant SDP regime emerges.

For detailed animations illustrating these three processes, we refer to the Supplementary Materials \cite{SM}.


\begin{figure}[t!]
  \includegraphics[width=\linewidth]{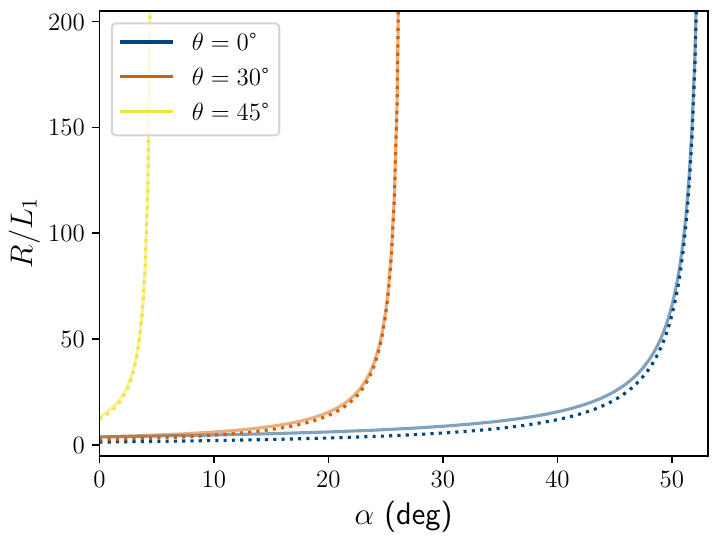}
  \caption{The growth of the Laplace radius $R$ with the opening angle $\alpha$ for systems in the DP condensation state. The results are presented for an aspect ratio of $a = 1.5$ and three different
    contact angles. Numerical results obtained from Eq.~(\ref{cc_dp}) (solid lines) confirm the asymptotic behavior $R \sim (\alpha_m - \alpha)^{-1}$, with the amplitude given by
    Eq.~(\ref{alpham_asym_dp2}) (dotted lines).} \label{Rlim_DP}
\end{figure}

\section{Asymptotic properties of condensation in open wedges} \label{asym}

This section examines the asymptotic behavior of condensation in open wedges in three limiting cases:\\
A) $\alpha\to\alpha_m$, B) $\alpha\to0$, and C) $a\gg1$.

\subsection{$\alpha\to\alpha_m$ limit}

We start by analyzing the divergence of the Laplace radius $R$ as the opening angle $\alpha$ approaches its maximum value, $\alpha_m(a,\theta)$,
allowing for condensation. Based on prior results, the analysis is restricted to two condensation regimes -- DP and SDP -- depending on the sum of
$\alpha$ and $\theta$.

\subsubsection{ $\alpha+\theta<\frac{\pi}{2}$: Divergence in DP condensation}

For $\alpha+\theta<\frac{\pi}{2}$, the system condenses in the DP regime as $\alpha$ approaches $\alpha_m$, which is determined from
Eq.~(\ref{amin_dp}). Recalling that for large $R$, the edge contact angles approach $\pi/2$ according to (\ref{theta1_lim}) and(\ref{theta2_lim}),
the condensation free-energy balance can be written as
\begin{equation}
2+a\tan\alpha-a(1+\sec\alpha)\cos\theta+\frac{L_1}{2R}\left(2+a\tan\alpha\right)a=0\,,
\end{equation}
which reduces to Eq.~(\ref{asym_dp}) for $\alpha=\alpha_m$ (when $R\to\infty$). Expanding this relation for $\alpha\to\alpha_m$ gives, to first
order:
\begin{multline}
  (\sin\alpha_m\cos\theta-1)a\sec^2\alpha_m(\alpha_m-\alpha)\\
  +\frac{L_1}{2R}\left(2+a\tan\alpha_m\right)a=0\,,
\end{multline}
which yields
\begin{equation}
R\sim f_{\rm DP}(a,\theta)(\alpha_m-\alpha)^{-1}\,, \label{alpham_asym_dp}
\end{equation}
with
\begin{equation}
f_{\rm DP}(a,\theta)=\frac{2+a\tan\alpha_m}{2\sec^2\alpha_m(1-\sin\alpha_m\cos\theta)}\,.  \label{alpham_asym_dp2}
\end{equation}

This result for the divergence of $R$ is borne by comparison with numerical solutions of the Kelvin equation for DP condensation, as shown in
Fig.~\ref{Rlim_DP}.

\subsubsection{ $\alpha+\theta>\frac{\pi}{2}$: Divergence in SDP condensation}

For $\alpha+\theta>\frac{\pi}{2}$, the system condenses in the SDP regime as $\alpha\to\alpha_m$, now determined by Eq.~(\ref{amin_sdp}). The
asymptotic behavior of $R$ as $\delta\alpha=\alpha_m-\alpha\to0$ follows from Eq.~(\ref{sd_therm}) but compared to the DP case the analysis is more
complex: i) At zeroth order in $1/R$, Eq.~(\ref{sd_therm}) is satisfied identically; ii) to first order, when $\theta_1=\pi/2-L_1/(2R)$ and
$\theta_2=\pi-\alpha_m-\theta-\frac{L_2}{R}\frac{\cos\alpha_m}{\sin\theta}+\delta\alpha$, Eq.~(\ref{sd_therm}) is satisfied due to
Eq.~(\ref{amin_sdp}); iii) at second order,  Eq.~(\ref{sd_therm}) becomes a quadratic form in both $1/R$ and $\delta\alpha$, which gives  the same
asymptotic scaling as in the DP case
\begin{equation}
R\sim f_{\rm SDP}(a,\theta)(\alpha_m-\alpha)^{-1}\,. \label{alpham_asym_sdp}
\end{equation}
but with a much more complicated amplitude:
\begin{widetext}
  \begin{multline}
    f_{\rm SDP}=\left\{\left[\cos\!\left(2 \alpha_m\right) - \cos\!\left(2 \alpha_m + 2 \theta\right) + 8 \cos\!\left(\alpha_m\right) + \cos\!\left(2 \theta\right) + 4 \sin\!\left(2 \alpha_m + \theta\right) + 8 \sin\!\left(\alpha_m + \theta\right) + 4 \sin\!\left(\theta\right) + 7\right]\vphantom{\frac{3}{2}}\right. \\
    \left. \left[\sin\!\left(\frac{1}{2} \alpha_m\right) - \sin\!\left(\frac{3}{2} \alpha_m\right)\right] \cot\!\left(\theta\right) \sec\!\left(\frac{1}{2} \alpha_m + \theta\right)\right\}\Bigg/\\
    \left[\vphantom{\frac{1}{1}}2 \cos\!\left(3 \alpha_m + \theta\right) - 8 \cos\!\left(\alpha_m + \theta\right) - 10 \cos\!\left(-\alpha_m + \theta\right) - \sin\!\left(3 \alpha_m\right) - 6 \sin\!\left(2 \alpha_m\right) - \sin\!\left(3 \alpha_m + 2 \theta\right) \right.\\
      \left. - 2 \sin\!\left(2 \alpha_m + 2 \theta\right) + 2 \sin\!\left(\alpha_m + 2 \theta\right) + 3 \sin\!\left(\alpha_m\right) - \sin\!\left(-\alpha_m + 2
      \theta\right) - 6 \sin\!\left(2 \theta\right)\vphantom{\frac{1}{1}}\right] \label{alpham_asym_sdp2}
  \end{multline}
\end{widetext}

\begin{figure}[t!]
  \includegraphics[width=\linewidth]{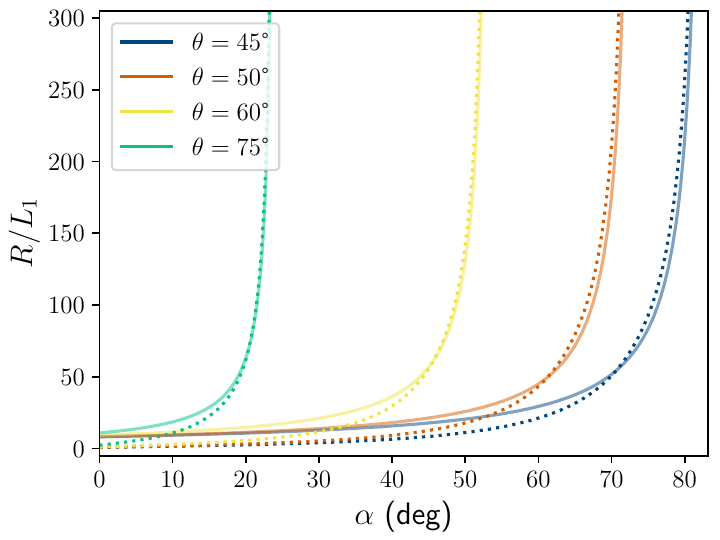}
  \caption{The growth of the Laplace radius $R$ with the opening angle $\alpha$ for systems in the SDP condensation state. The results are presented for an aspect ratio of $a = 20$ and four different
    contact angles. Numerical results obtained from Eq.~(\ref{sd_therm}) (solid lines) confirm the asymptotic behavior $R \sim (\alpha_m - \alpha)^{-1}$, with the amplitude given by
    Eq.~(\ref{alpham_asym_sdp}) (dotted lines).} \label{Rlim_SDP}
\end{figure}


This asymptotic result is verified by numerical solutions of Eq.~(\ref{sd_therm}), with the comparison shown in Fig.~\ref{Rlim_SDP}.


\begin{figure}[b!]
  \includegraphics[width=\linewidth]{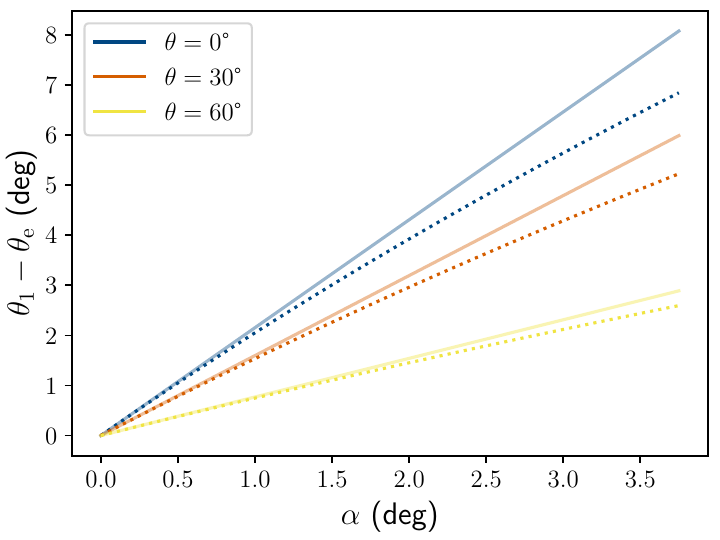}
  \caption{Numerical confirmation (for $a=3$) of the asymptotic prediction (\ref{theta_alpha}), which states that the edge contact angle $\theta_1$
    converges linearly to the edge contact angle $\theta_e$ (pertinent to a finite slit) as $\alpha \to 0$. The dotted lines represent the numerical
    solutions obtained from Eq.~(\ref{cc_dp}), while the solid lines correspond to the analytic result (\ref{theta_alpha}). In all cases,
    $\theta_1(\alpha)$ approaches $\theta_e$ from above, as expected.} \label{theta_to_0}
\end{figure}

\subsection{$\alpha\to0$ limit}

In the limit where the opening angle $\alpha$ vanishes, the wedge approaches a near-parallel wall system resembling a finite slit. In this geometry, the only possible condensation regime is the DP
state. The free-energy balance, Eq.~(\ref{cc_dp}), simplifies in this limit to
\begin{multline}
  \frac{2L_1+\alpha H}{2R}H+R(\pi-\theta_1-\theta_2)-2H\cos\theta\\
  +\frac{L_1}{2}\sin\theta_1+\frac{L_1}{2}\sin\theta_2+\frac{\alpha H}{2}\sin\theta_2=0\,, \label{R2_small_alpha}
\end{multline}
where Eqs.~(\ref{cos_theta1}), (\ref{dp_theta2}), and the approximation $L_2\approx L_1+\alpha H$ have been used.  Combining these relations further
implies that
\begin{equation}
\cos\theta_2=\cos\theta_1+\frac{\alpha}{2}\frac{H}{R}\,,
\end{equation}
from which we obtain for the difference between the two edge contact angles
\begin{equation}
\theta_1-\theta_2=a\alpha\cot\theta_1\,, \;\;\;(\alpha\to0)\,.\label{delta_theta}
\end{equation}
Using this expression, $\theta_2$ can be eliminated from the free-energy balance equation, leading to
\begin{multline}
 \hspace{-\multlinegap}
  2a\cos\theta_1+\alpha a^2\cos\theta_1+\sec\theta_1\left(\frac{\pi}{2}-\theta_1\right)+\frac{\alpha a}{2}\sec\theta_1\cot\theta_1\\
  -2a\cos\theta+\sin\theta_1-\frac{\alpha a}{2}\cos\theta_1\cot\theta_1+\frac{\alpha a}{2}\sin\theta_1=0\,, \label{R2_small_alpha2}
\end{multline}
which  for $\alpha=0$ reduces to
\begin{equation}
\cos\theta_1=\cos\theta-\frac{1}{2a}\left[\sin\theta_1+\sec\theta_1\left(\frac{\pi}{2}-\theta_1\right)\right]\,, \;\;\;(\alpha=0)\,,\label{finite}
\end{equation}
recovering Kelvin's equation for finite slits (cf. Eq.~(\ref{theta_e})).

For small $\alpha$, the $\alpha$ dependence of $\theta_1$ can be approximated as
\begin{equation}
\theta_1(\alpha)\approx\theta_e+\lambda(a,\theta)\alpha\;\;\;(\alpha\to0)\,, \label{theta_alpha}
\end{equation}
where $\theta_e=\theta_1(0)$ (the edge contact angle for a finite slit) and
\begin{equation}
\lambda(a,\theta)=\frac{(a^2\cos\theta_e+a\sin\theta_e)\cot\theta_e}
{2a\cos\theta_e+\sin\theta_e+\sec\theta_e\left(\theta_e-\frac{\pi}{2}\right)}\,.
\end{equation}
 This approximation simplifies further for systems with $a\gg1$ (in which case the DP regime is stable only for very small values of $\alpha$)
\begin{equation}
\theta_1(\alpha)\approx\theta_e+\frac{a}{2}\cot\theta_e\alpha\;\;\;(\alpha\to0)\,,
\end{equation}
showing a linear approach of $\theta_1$ to $\theta_e$ from above as $\alpha\to0$, which upon substituting to Eq.~(\ref{delta_theta}) reveals the
symmetric behavior for $\theta_2$
\begin{equation}
\theta_2(\alpha)\approx\theta_e-\frac{a}{2}\cot\theta_e\alpha\;\;\;(\alpha\to0)\,,
\end{equation}
approaching $\theta_e$ from below.

The asymptotically linear convergence of $\theta_1$ to $\theta_e$ in the limit of $\alpha\to0$ is confirmed by the numerical solution obtained from
Eq.~(\ref{cc_dp}), as shown in Fig.~\ref{theta_to_0}.

\begin{figure}[t!]
  \includegraphics[width=0.75\columnwidth]{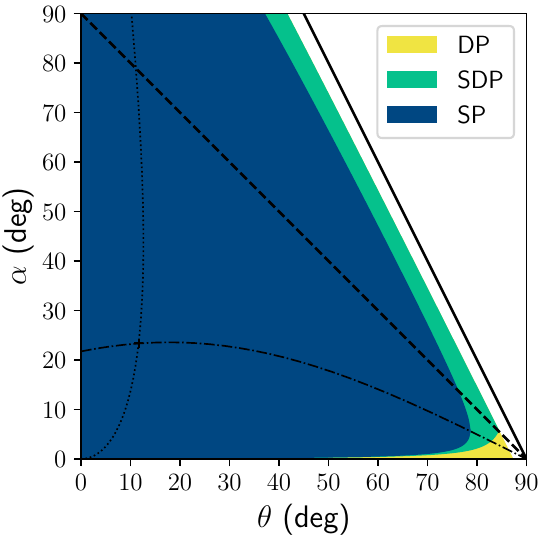}
  \caption{Phase diagram in the $\alpha$-$\theta$ plane for a very long system with aspect ratio $a = 20$. The condensed region is predominantly of SP
    type, except for small values of $\alpha$, where the condensation is of DP type, and near the condensation limit $\alpha + 2\theta = \pi/2$, where
    the condensation is of SDP type.} \label{pd_20}
\end{figure}

\subsection{$a\gg1$}

Finally, we examine the impact of extending the walls by considering open wedges with very large aspect ratios. In this case, condensation is
predominantly of an SP type, as demonstrated in Fig.~\ref{pd_20}. This is because the free-energy cost associated with the right meniscus in DP and
SDP configurations becomes significant as $a$, and consequently $L_2$, is large, except for two cases:
\begin{enumerate}
\item[i)] Near the condensation limit, $\alpha+2\theta=\pi/2$, the menisci become nearly flat, stabilizing the SDP state.
\item[ii)] For sufficiently small values of $\alpha$, the system inevitably adopts a DP configuration.
\end{enumerate}

Simplifications in the large $a$ limit are thus only expected in these  two specific scenarios, as the wall's extension does not affect the SP
configuration. While the SDP condensation near the condensation limit has already been analyzed in part {\bf B}, for DP condensation we can also take
advantage of our previous findings, exploiting the fact that only small values of $\alpha$ are relevant. Rearranging  Eq.~(\ref{R2_small_alpha2}), we
obtain
\begin{multline}
  \frac{a}{\tilde{R}}\left(1+\frac{\alpha a}{2}\right)+2\tilde{R}\left(\frac{\pi}{2}-\theta_1\right)+\alpha a\frac{\sqrt{4\tilde{R}^2-1}}{2\tilde{R}}\\
  -2a\cos\theta+\frac{\sqrt{4\tilde{R}^2-1}}{2\tilde{R}}=0\,,   \label{ap_large_a}
\end{multline}
where we abbreviated $\tilde{R}=R/L_1$. As shown in Appendix \ref{app_a}, the solution of Eq.~(\ref{ap_large_a}) in the regime of $a\gg1$ and
$\alpha\ll1$ is
\begin{equation}
\tilde{R}\approx\frac{a\left(1+\frac{\alpha a}{2}\right)}{2a\cos\theta-\alpha a-2} \label{R_asym}\,.
\end{equation}

\begin{figure}[t!]
  \includegraphics[width=\linewidth]{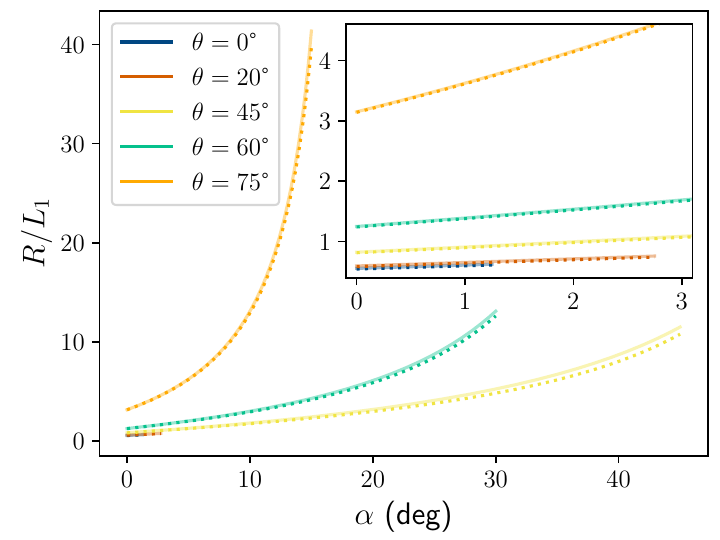}
  \caption{Numerical verification of the approximate relation for the location of DP condensation (\ref{R_asym}) for large values of $a$. The comparison is made between the approximation and the
    numerical results obtained from (\ref{cc_dp}) for $a=6$ and various contact angles.} \label{Rapp}
\end{figure}

In Fig.~\ref{Rapp}, we show a very good agreement between this prediction and the numerical solution of Eq.~(\ref{cc_dp}).

\section{Summary}

In this work, we studied condensation in open wedges embedded into a gas reservoir. On a macroscopic level, the condensation is fully described by
the three wedge parameters---the aspect ratio $a=H/L_1$, the opening angle $\alpha$, and the Young's contact angle $\theta$---and the condensed
states by the menisci delineating the condensed phase. We demonstrated that condensation can manifest in one of three distinct configurations: SP
(single pinning), SDP (semi-double pinning), or DP (double pinning), which differ by the location of the right meniscus near the wider opening:
\begin{itemize}
  \item \emph{SP:} The meniscus meets both walls inside the wedge.

  \item \emph{DP:} The meniscus connects the walls at their edges.

  \item \emph{SDP:} An intermediate state where the meniscus is pinned to one edge but is free to slide along the opposite wall.

\end{itemize}

We derived the conditions under which each condensation type occurs, the transitions between them, and the asymptotic behavior in specific limiting
cases. Our key findings are summarized as follows:
\begin{enumerate}
  \item General conditions for condensation:
         \begin{enumerate}

          \item  Capillary condensation can only occur if  $\alpha/2+\theta<\pi/2$.

          \item The DP configuration requires an additional constraint $\alpha+\theta<\pi/2$.

        \end{enumerate}

  \item In the limit of $p\to p_{\rm sat}^-$ ($\delta p\to0^+$), condensation is restricted to DP or SDP configurations. Specifically:
       \begin{enumerate}

          \item \emph{DP condensation}: Occurs if  $\sin\alpha<\cos\theta$ and $a>a^{\rm DP}(\alpha,\theta)$, where $a^{\rm DP}$ is given by Eq.~(\ref{amin_dp}).

          \item \emph{SDP condensation}: Occurs if $\sin\alpha>\cos\theta$ and $a>a^{\rm SDP}(\alpha,\theta)$, where $a^{\rm SDP}$ is given by Eq.~(\ref{amin_sdp}).

          \item If neither condition is met, the system remains in a low-density state.

        \end{enumerate}

  \item The phase behavior dependence on the system size (aspect ratio):
         \begin{enumerate}

          \item Condensation may occur only if $a>1$.

          \item $1<a<2$ (``short systems''): Only DP condensation is possible.

          \item   $2<a<a^{\rm SP}$ (``moderately long systems''), with $a^{\rm SP}\approx 4.53$:

                \begin{itemize}

                  \item If $a\cos\theta<2$: Condensation is of DP type.

                  \item If $a\cos\theta\ge 2$: DP occurs for $\alpha<\tilde{\alpha}(\theta)$, while SDP occurs for $\alpha>\tilde{\alpha}(\theta)$; $\tilde{\alpha}(\theta)$ is given by Eq.~(\ref{dp_sdp}).

                \end{itemize}

          \item $a>a^{\rm SP}$ (``long systems''): All three types (SP, SDP, DP)  can occur.
          The SP region in the global phase diagram grows with $a$ and for $a\gtrapprox 8.56$  approaches the limit $\alpha=\pi/2$.

        \end{enumerate}

  \item For given $\theta$ and $a$, there exists a maximal opening angle $\alpha_m$, beyond which condensation cannot occur. As $\alpha\to\alpha_m$, the
        Laplace radius diverges as $R\sim (\alpha_m-\alpha)^{-1}$.

  \item As $\alpha\to0$, the system resembles a finite slit, with DP being the only condensation type. The edge contact angles $\theta_1$ and $\theta_2$ approach the slit contact angle $\theta_e$,
        with $\theta_1$ converging from above and $\theta_2$ from below.
        For long slits: $\theta_{1,2}\approx\theta_e\pm a/2\cot\theta_e\alpha$.

  \item For large aspect ratios, $a\gg1$,  condensation is predominantly of SP type. DP condensation is limited to small $\alpha$ and
        follows the asymptotic relation Eq.~(\ref{R_asym}).
        SDP persists only near the condensation limit ($\alpha+2\theta=\pi$), where the Laplace radius scales as $R\sim a$.

  \item Continuous transitions occur between SP and SDP (at pressure given by Eq.~(\ref{p_sp_sdp})) and between SDP and DP
       (at pressure given by Eq.~(\ref{p_sdp_dp})).
        The transitions are second-order for partially wet walls and third-order for completely wet walls.

\end{enumerate}

In summary, we developed a macroscopic framework for describing condensation in open wedges, revealing a rich phenomenology of phase transitions arising from the interplay of geometric and thermodynamic parameters.
We systematically analyzed the conditions for specific transitions and their asymptotic behavior, providing a comprehensive understanding of the model's phase diagram.
We identified three distinct types of first-order condensation transitions and two continuous transitions between different condensation states.
Furthermore, we derived Kelvin-like equations governing each condensation type and determined the location and nature of the continuous transitions.

There are several directions in which the present study could be extended. 
A particularly important extension would be the incorporation of a more microscopic perspective. 
This would introduce additional length scales, leading to a phase behavior that depends on $H$ and $L_1$ independently. 
A microscopic treatment is especially relevant for molecularly narrow wedges, where the influence of thick wetting layers and packing effects are particularly significant and which could be explored using a classical density functional theory.
Furthermore, it would be interesting to investigate the role of wetting layers in the possible rounding of the phase transitions and to examine the impact of intermolecular forces and thermal interfacial fluctuations by formulating an appropriate scaling theory.

Another potential extension concerns the model of the confining walls. For example, one could explore the effects
of wall roughness and modifications to phase behavior arising from variations in chemical composition. Additionally,
relaxing the assumption of translational symmetry in favor of axial symmetry could lead to new insights. Moreover,
identifying alternative wall geometries that exhibit phase transitions that can be related to those studied here may
reveal deeper underlying symmetries, contributing to a broader understanding of interfacial phenomena.

Finally, an intriguing direction would be to extend the present study beyond equilibrium by e.g. investigating
confined self-propelled particles as a model for active matter. We intend to address some of these extensions in our
future work.

\appendix

\section{Derivation of formula (\ref{sd_theta2})} \label{app_sdp}

\begin{figure}[h]
  \includegraphics[width=\linewidth]{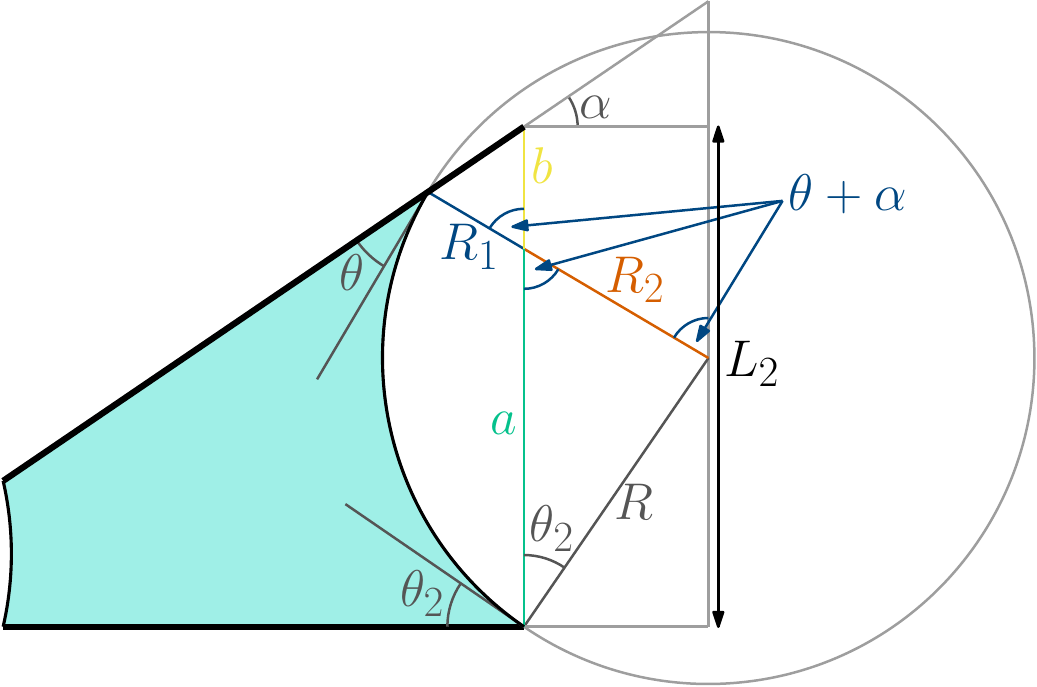}
  \caption{Sketch of an SDP configuration.} \label{fig_app_a}
\end{figure}

To derive Eq.~(\ref{sd_theta2}), we use basic trigonometry based on the scheme shown in Fig.~\ref{fig_app_a}. As indicated, the lengths $L_2$ and $R$
are decomposed as follows $L_2=a+b$ and $R=R_1+R_2$. Applying  the sine law to the lower triangle, we have
\begin{equation}
\frac{a}{\sin(\pi-\theta_2-\theta-\alpha)}=\frac{R}{\sin(\theta+\alpha)}\,,
\end{equation}
implying
\begin{equation}
a=R\frac{\sin(\theta_2+\theta+\alpha)}{\sin(\theta+\alpha)}\,.
\end{equation}
Additionally
\begin{equation}
\frac{R_2}{\sin\theta_2}=\frac{R}{\sin(\theta+\alpha)}\,,
\end{equation}
leading to
\begin{equation}
R_2=R\frac{\sin\theta_2}{\sin(\theta+\alpha)}\,.
\end{equation}

Applying the sine law to the upper triangle, we get
\begin{equation}
\frac{b}{\sin(\pi/2-\theta)}=\frac{R_1}{\sin(\pi/2-\alpha)}\,,
\end{equation}
hence
\begin{equation}
b=R_1\frac{\cos\theta}{\cos\alpha}\,.
\end{equation}
Substituting for $R_1=R\left(1-\frac{\sin\theta_2}{\sin(\theta+\alpha)}\right)$, we thus find
\begin{equation}
b=R\left(1-\frac{\sin\theta_2}{\sin(\theta+\alpha)}\right)\frac{\cos\theta}{\cos\alpha}\,.
\end{equation}

Now, combining $a$ and $b$, we determine the ratio $L_2/R$:
\begin{multline}
  \frac{L_2}{R}=\frac{a+b}{R}\\
  =\frac{\sin(\theta_2+\theta+\alpha)}{\sin(\theta+\alpha)}
  +\left(1-\frac{\sin\theta_2}{\sin(\theta+\alpha)}\right)\frac{\cos\theta}{\cos\alpha}\\
=\frac{\cos\theta}{\cos\alpha}+\cos\theta_2+\sin\theta_2\frac{\cos(\theta+\alpha)\cos\alpha-\cos\theta}{\sin(\theta+\alpha)\cos\alpha}\\
=\frac{\cos\theta}{\cos\alpha}+\cos\theta_2-\sin\theta_2\frac{\cos\theta\sin^2\alpha+\sin\theta\sin\alpha\cos\alpha}{\sin(\theta+\alpha)\cos\alpha}\\
=\frac{\cos\theta}{\cos\alpha}+\cos\theta_2-\sin\theta_2\tan\alpha\\
  =\frac{\cos\theta+\cos(\theta_2+\alpha)}{\cos\alpha}\,,
\end{multline}
which completes the derivation of Eq.~(\ref{sd_theta2}).

\section{Derivation of formula (\ref{R_asym})} \label{app_a}

To derive formula (\ref{R_asym}), we analyze Eq.~(\ref{ap_large_a}) under the asymptotic conditions $\alpha \ll 1$ and $a \gg 1$. The analysis is
divided into three cases based on the relative magnitudes of $a$ and $\alpha^{-1}$. Generally, the penultimate term in Eq.~(\ref{ap_large_a})
dominates, except when $\theta$ is close to $\pi/2$, requiring separate consideration.\\

\begin{enumerate}

  \item Case $a\alpha\gg1$:

        In this limit, for sufficiently small $\theta$, Eq.~(\ref{ap_large_a}) balances only if $\tilde{R}$ is of the order of $\alpha a$, leading to:
        \begin{equation}
        \tilde{R}=\frac{\alpha a}{4\cos\theta}\,. \label{R1}
        \end{equation}

        When $\theta$ approaches $\pi/2$, we have $\cos \theta \approx \alpha$ and $\theta_1 \approx \pi/2 - 1/(2 \tilde{R})$. In this case, $\tilde{R}$
        scales as $a$, and we obtain:
        \begin{equation}
        \tilde{R}=\frac{\alpha a}{4\cos\theta-2\alpha}\,. \label{R2}
        \end{equation}
        This result generalizes Eq.~(\ref{R1}), making it applicable for all $\theta$.

  \item Case $a\alpha\ll1$:

        In this scenario, for $\theta$ sufficiently far from $\pi/2$, Eq.~(\ref{ap_large_a}) balances only if $\tilde{R} \ll a$, giving
        \begin{equation}
        \tilde{R}=\frac{1}{2\cos\theta}\,. \label{R3}
        \end{equation}
        When $\theta \approx \pi/2$, $\tilde{R} \sim a$, and we find
        \begin{equation}
        \tilde{R}=\frac{a}{2a\cos\theta-2}\,,\label{R4}
        \end{equation}
        which is an extension of Eq.~(\ref{R3}).

  \item Case $a\alpha\sim1$:

        In this intermediate regime, for sufficiently small $\theta$, the penultimate term in Eq.~(\ref{ap_large_a}) balances with the first term, resulting
        in
        \begin{equation}
        \tilde{R}=\frac{1+\frac{\alpha a}{2}}{2\cos\theta} \label{R5}\,.
        \end{equation}
        When $\theta \approx \pi/2$, all terms in Eq.~(\ref{ap_large_a}) become relevant, and $\tilde{R}$ scales with $a$
        \begin{equation}
        \tilde{R}=\frac{a\left(1+\frac{\alpha a}{2}\right)}{2a\cos\theta-\alpha a-2} \label{R6}\,.
        \end{equation}

This result represents the most general asymptotic solution of Eq.~(\ref{ap_large_a}) and describes the crossover between Eqs.~(\ref{R2}) and
(\ref{R4}). Hence, Eq.~(\ref{R6}) provides the sought expression for the behavior of $\tilde{R}$ in the asymptotic regime $a \gg 1$ and $\alpha \ll
1$.

\end{enumerate}

\end{document}